\documentclass[usenatbib,referee]{biom}
\usepackage{amsmath}
\usepackage{bbm}
\usepackage{setspace}
\usepackage{xcolor}
\usepackage{float}
\usepackage{amssymb}
\usepackage{graphicx}
\usepackage{natbib} 
\usepackage{url} 
\usepackage{hyperref}
\usepackage{algorithm} 
\usepackage{algpseudocode} 

\def\bSig\mathbf{\Sigma}

\title[]{Combining Mixed Effects Hidden Markov Models with Latent Alternating Recurrent Event Processes to Model Diurnal Active-Rest Cycles}

\author
{Benny Ren\emailx{bennyren@pennmedicine.upenn.edu} \\
Department of Biostatistics, Epidemiology, and Informatics, \\ 
University of Pennsylvania, Philadelphia, U.S.A.
\and 
Ian Barnett\emailx{ibarnett@pennmedicine.upenn.edu} \\
Department of Biostatistics, Epidemiology, and Informatics, \\ 
University of Pennsylvania, Philadelphia, U.S.A.}

\begin{document}


\date{}



\pagerange{\pageref{firstpage}--\pageref{lastpage}} 
\volume{}
\pubyear{}
\artmonth{}


\doi{}


\label{firstpage}


\begin{abstract}
Data collected from wearable devices and smartphones can shed light on an individual's pattern of behavioral and circadian routine. Phone use can be modeled as alternating event process, between the state of active use and the state of being idle. Markov chains and alternating recurrent event models are commonly used to model state transitions in cases such as these, and the incorporation of random effects can be used to introduce diurnal effects. While state labels can be derived prior to modeling dynamics, this approach omits informative regression covariates that can influence state memberships. We instead propose an alternating recurrent event proportional hazards (PH) regression to model the transitions between latent states.  We propose an Expectation-Maximization (EM) algorithm for imputing latent state labels and estimating regression parameters. We show that our E-step simplifies to the hidden Markov model (HMM) forward-backward algorithm, allowing us to recover a HMM with logistic regression transition probabilities. In addition, we show that PH modeling of discrete-time transitions implicitly penalizes the logistic regression likelihood and results in shrinkage estimators for the relative risk. We derive asymptotic distributions for our model parameter estimates and compare our approach against competing methods through simulation as well as in a digital phenotyping study that followed smartphone use in a cohort of adolescents with mood disorders.
\end{abstract}

%



\maketitle


%

\section{Introduction}
\label{intro}

Diurnal and circadian rhythm studies often model physiological processes as periodic cycles, such as a person's active and rest cycle. Sleep and diurnal rhythm are essential components of many circadian physiological processes with a clear time-of-day effect on active and rest cycles \citep{lagona2014latent,morris2012circadian}. While classification of physiological processes is an ongoing area of research, in many instances, processes can be discretized into a few state categories such as active and rest state labels. 
Here we consider this problem of estimating an individual's cycles between active and rest states in a mobile health (mHealth) setting based on wearable device or smartphone sensor data. If the true state labels are known, then a Markov chain can be used to model state transitions over time, otherwise a hidden Markov model (HMM) can be used to simultaneously perform classification and state transition estimation \citep{langrock2013combining}.
In addition, HMMs have been extended to incorporate time-of-day effects as periodicity or seasonality using random effects \citep{stoner2020advanced,holsclaw2017bayesian,bartolucci2015discrete}. Continuous-time hidden Markov models (CT-HMM) have also been used for state classification in similar contexts but have difficulty accounting for random effects \citep{bartolucci2019shared,liu2015efficient,bureau2003applications,jackson2003multistate}. Most mixed effects HMM estimation procedures estimate logistic regression for discrete-time processes and do not account for time between states \citep{maruotti2012mixed,altman2007mixed}.

It is important to note that active and rest state transitions are ergodic processes and the sojourn time between these states can be modeled with a proportional hazards (PH) regression in both directions, active-to-rest and rest-to-active. These two directions of transitions can be viewed as an alternating recurrent event PH model \citep{krol2015semimarkov,wang2020penalized,shinohara2018alternating}. Already, hazard rates have been incorporated into continuous-time Markov chains (CTMC) to model sojourn times \citep{hubbard2016using}. However, the ability for alternating recurrent event processes to accurately model sojourn times complements HMMs and provides many useful properties in addition to being computationally scalable. This novel model can be viewed as a latent state analog of an alternating recurrent event process \citep{wang2020penalized,krol2015semimarkov}. Mainly, if the underlying data generating process is a continuous-time process, then PH models are a more appropriate modeling choice \citep{abbott1985logistic,ingram1989empirical}. If the data generating process involves discrete-time transitions, we showed that PH modeling penalizes the logistic regression likelihood, inducing shrinkage during estimation. 

We propose an approach that takes advantage of the strengths of both HMMs and alternating recurrent event models to jointly estimate latent states while simultaneously providing flexible modeling of sojourn times. Our Expectation-Maximization (EM) algorithm imputes latent active and rest state labels while modeling state transitions with an alternating recurrent event process using exponential PH regressions \citep{dempster1977maximum}. Informative regression covariates, often omitted in state labeling, are incorporated into the latent state imputation using the EM algorithm. Under the EM algorithm, we show that the E-step in this case simplifies to imputations involving a HMM forward-backward algorithm where state transition probabilities are defined as logistic or multinomial regression probabilities \citep{baum1970maximization,altman2007mixed}. We also show that the M-step reduces to fitting independent PH models weighted by E-step imputations allowing for a potentially large number of latent states, as well as, providing a means to obtain large sample theory inference. Our EM approach involving PH models provides a scaleable M-step while returning multinomial regression transition probabilities commonly found in HMMs \citep{maruotti2012mixed,holsclaw2017bayesian,altman2007mixed}. Furthermore, we show that applying PH models to discrete-time transitions, implicitly penalizes logistic regression to shrink the transition probability matrices of HMMs towards the identity matrix and mitigates overfitting in many practical settings. As a result, the PH models favors processes with a low incidence of state transitions such as diurnal cycles where we expect few state transitions within a 24h period.

We apply this approach to estimate active-rest diurnal cycles in a sample of patients with affective disorders using their passively collected smartphone sensor data, namely through the accelerometer, screen on/off data of patient smartphones and time-of-day random intercepts. We are able to quantify the strength of a patient's routine by representing the magnitude of time-of-day random intercepts as the regularity of a patient's diurnal rhythm. This quantification of the strength of routine can be correlated with a myriad of relevant clinical outcomes, as the regularity of diurnal rhythms plays an important role in psychopathology with past studies having shown associations between irregular rhythms and adverse health outcomes \citep{monk1990social,monk1991social}.
In addition, we fit a population level HMM to study effects of individual specific covariates on state transitions.
\section{Data and Methods}
\label{methods}


Our data consist of $i \in \{ 1,\dots, I \}$ individuals, with each individual $i$ having a sequence $j \in \{ 1, \dots, n_i \}$ of covariates to be modeled with separate hidden Markov models. The HMM of the $i$th individual has $n_i+1$ sequences of active or rest states $\mathbf{A}_i = \left\{ A(t_{i0}),\dots,A(t_{ij}),\dots,A(t_{in_i}) \right\}$, where hourly time-stamps $t_{j}$ are increasing in $j$. In our example, we denote active and rest states as $A(t_{ij})=1$ and $A(t_{ij})=2$ respectively, with an outline of our HMM in Web Figure 1. Within each sequence, we define event times for $n_i$ state transitions as $\Delta(t_{ij}) = t_{ij} - t_{i(j-1)}$, which follow from an exponential distribution and can be fitted with a PH regression. Linking multiple exponential event time processes results in a recurrent event model. Furthermore, recurrent exponential PH models are analogous with a non-homogeneous Poisson processes which retains independent increments, allowing us to chain multiple transitions together. 

The covariates used in the exponential PH regression are mean acceleration magnitudes (Euclidean norm) from the preceding hour evaluated at $n_{i}$ transitions and are outlined in Web Figure 1. We denote the intercept and covariates as $\mathbf{X}^\top_{i} = \left[ \mathbf{x}(t_{i1}),\mathbf{x}(t_{i2}),\dots,\mathbf{x}(t_{i n_{i}}) \right] \in \mathbb{R}^{p \times n_i}$. We make an ergodic state transition assumption where states will inevitably communicate with each other, i.e., the active-to-rest and rest-to-active transitions will eventually occur. This allows the survival function to capture the likelihood contribution of when a state transition did not occur, meaning the transition will occur at some future time. Because we do not have the true state labels $\mathbf{A} = \{\mathbf{A}_1,\dots \mathbf{A}_I \}$, we must rely on state dependent observations. We use screen-on counts for each time-stamp $\mathbf{y}_{i} = \left\{ Y(t_{i0}),Y(t_{i1}),\dots,Y(t_{i n_i }) \right\}$ as observations from state dependent distributions $Y(t_{ij}) | \left\{ A({t_{ij}}) = s \right\} \sim \text{Poisson}( \mu_s )$, where $s \in \{1,2\}$, $\mu_s$ are state specific parameters and we expect $\mu_2 \approx 0$ for the rest state.

\subsection{Alternating Recurrent Event PH Model}
\label{PHHMM}

In our two state setting, rates of transition from state $s$ to the other state are defined as $\lambda_s(t_{ij}) = \exp \left( \mathbf{x}^\top(t_{ij}) \boldsymbol{\beta}_s \right)$. For example, $\lambda_1(t_{ij})$ denotes the rate of transition from state 1-to-2 (active-to-rest). Alternating recurrent event PH models often need to account for the longitudinal nature of the data, i.e., repeated measurements. Mixed effects or frailties can be used to account for the recurrent nature of the data \citep{wang2020penalized,mcgilchrist1991regression}. Modifying the standard exponential PH model with a shared log-normal frailties or normal random intercepts, state transition hazards become 
$ \lambda_s(t_{ij}) = \exp \left( \eta_s(t_{ij}) \right) = \exp \left( \mathbf{x}^\top(t_{ij}) \boldsymbol{\beta}_s + \mathbf{z}^\top(t_{ij}) \mathbf{b}_s \right)$, where $\mathbf{b}_s \sim \mathrm{N}(\mathbf{0}_{24}, \sigma^2_s \mathbf{I}_{24} )$. Here $\mathbf{z}(t_{ij})$ are 24 hour-of-day indicators, one-hot vectors designed to toggle the appropriate random intercepts within $\{ \mathbf{b}_1,\mathbf{b}_2 \}$.

A HMM for individual $i$ is given by the complete data likelihood
\begin{equation} \label{eq1}
\begin{array}{rl}
L ( \boldsymbol{\beta}_1, \mathbf{b}_1, \sigma^2_1, \boldsymbol{\beta}_2, \mathbf{b}_2, \sigma^2_2, \mu_1, \mu_2 | \mathbf{A}_i ) &= 
\left\{ \prod_{s=1}^2 L(\boldsymbol{\beta}_s, \sigma^2_s, \mathbf{b}_s |\mathbf{A}_i ) \right\} L ( \mu_1, \mu_2 | \mathbf{A}_i ) \\
&= \left\{ \prod_{s=1}^2 L(\boldsymbol{\beta}_s | \mathbf{A}_i, \mathbf{b}_s ) f ( \mathbf{b}_s | \sigma^2_s ) \right\} L ( \mu_1, \mu_2 | \mathbf{A}_i ) 
\end{array}
\end{equation}
where $\mathbf{A}_i$ are the true the state labels. The PH likelihoods for state transitions are given as
$$
L(\boldsymbol{\beta}_s | \mathbf{A}_i, \mathbf{b}_s ) f ( \mathbf{b}_s | \sigma^2_s ) = \left[ \prod^{n_{i}}_{j=1} \left\{ f ( \Delta(t_{ij}) | \lambda_s(t_{ij}) ) \right\}^{d_s(t_{ij})} \left\{ S ( \Delta(t_{ij}) | \lambda_s(t_{ij}) ) \right\}^{c_s(t_{ij})} \right] f ( \mathbf{b}_s | \sigma^2_s )
$$
where $f ( \Delta(t_{ij}) | \lambda_s(t_{ij}) ) = \lambda_s(t_{ij}) \exp \left( -\lambda_s(t_{ij}) \Delta(t_{ij}) \right)$ and $S ( \Delta(t_{ij}) | \lambda_s(t_{ij}) ) = \exp \left( -\lambda_s(t_{ij}) \Delta(t_{ij}) \right)$ are derived from the exponential distribution. Note that $\prod_{s=1}^2 L(\boldsymbol{\beta}_s, \sigma^2_s, \mathbf{b}_s |\mathbf{A}_i ) f ( \mathbf{b}_s | \sigma^2_s )$ is the likelihood of an alternating event process \citep{krol2015semimarkov,wang2020penalized}. We denote indicators for state 1-to-2 transitions $d_1(t_{ij}) = \mathbb{I} \left[ A(t_{i(j-1)})=1, A(t_{ij}) = 2 \right]$, as $\mathbf{d}_1 = \left\{ d_1(t_{11}), d_1(t_{12}), \dots, d_1(t_{ij}), \dots \right\}$ and 2-to-1 transitions as $\mathbf{d}_2$. We interpret failure to transition out of state 1, $c_1(t_{ij}) = \mathbb{I} \left[ A(t_{i(j-1)})=1, A(t_{ij}) = 1 \right]$ as censoring, denoted as $\mathbf{c}_1 = \left\{ c_1(t_{11}), c_1(t_{12}), \dots, c_1(t_{ij}), \dots \right\}$ and we similarly define $\mathbf{c}_2$. The screen-on count state conditional Poisson likelihood is given as $ \begin{array}{c}
L \left( \mu_1, \mu_2 | \mathbf{A}_i \right) = \prod^{n_{i}}_{j=0} \prod^{2}_{s=1} f \left( y(t_{ij}) | \mu_s \right)^{u_s(t_{ij})}
\end{array} $ where state memberships $\mathbf{u}$, are denoted as indicators $u_s(t_{ij}) = \mathbb{I} \left[ A(t_{ij}) = s \right]$. Since the true labels are unknown, $\mathbf{d}_s$, $\mathbf{c}_s$, and $\mathbf{u}$ are latent variables and \eqref{eq1} becomes a mixture model.

\subsection{EM Algorithm for PH Regression and HMM Parameters}
\label{est}
The log-likelihood of \eqref{eq1} are linear functions of latent variables $\mathbf{d}_s$, $\mathbf{c}_s$, and $\mathbf{u}$, lending our optimization approach to an EM algorithm \citep{dempster1977maximum}. Through the EM algorithm, indicators $\mathbf{d}_s$, $\mathbf{c}_s$, and $\mathbf{u}$ are imputed as continuous probabilities, in the process of obtaining maximum likelihood estimates (MLEs). As a result, the alternating recurrent event exponential PH model reduces to two weighted frailty models. We denote PH model weights as $\mathbf{w}(t_{ij}) = \left\{ {c}_1(t_{ij}), {d}_1(t_{ij}), {c}_2(t_{ij}), {d}_2(t_{ij}) \right\}$, which belong to a 4-dimensional probability simplex, i.e., values are non-negative and $\| \mathbf{w}(t_{ij}) \|_1 = 1$. Poisson mixture model weights $\mathbf{u}(t_{ij}) = \left\{ {u}_1(t_{ij}), u_2(t_{ij}) \right\}$ belong to a 2-dimensional probability simplex. 
Our EM algorithm iteratively estimates the weights $\mathbf{w}(t_{ij})$ and $\mathbf{u}(t_{ij})$ using the forward-backward algorithm of \cite{baum1970maximization} and $\left\{ \boldsymbol{\beta}_1, \mathbf{b}_1, \sigma^2_1, \boldsymbol{\beta}_2, \mathbf{b}_2, \sigma^2_2 \right\}$ using survival modeling. While the complete data likelihood is written as an alternating event processes, we see an equivalence with logistic regression transition probabilities in the E-step calculations, effectively retooling alternating event processes to fit HMMs to data that have heterogeneous event times transitions.

\subsubsection{E-step}
\label{EStep}
In the E-step, we derive the expectation of $\mathbf{w}(t_{ij})$ conditional on model parameters and observed data in the $i$th HMM denoted by $\mathbf{y}_{i}$ and $\mathbf{X}_{i}$, as
$$ 
\begin{array}{rl}
\mathbb{E} \left[ {d}_1(t_{ij}) | \mathbf{y}_{i}, \mathbf{X}_{i}, {\Theta}_{i} \right] &= 
\text{Pr} \left( A(t_{i(j-1)})=1, A(t_{ij}) = 2 | \mathbf{y}_{i}, \mathbf{X}_{i}, {\Theta}_{i} \right) \\
&= \left( \frac{ {\alpha}_1 (t_{i(j-1)}) {\nu}_2 (t_{ij})}{ \text{Pr}(\mathbf{Y}_{i} =\mathbf{y}_{i} | \mathbf{X}_{i}, {\Theta}_{i}) } \right) \frac{f ( \Delta(t_{ij}) | \lambda_1(t_{ij}) ) }{f ( \Delta(t_{ij}) | \lambda_1(t_{ij}) ) + S ( \Delta(t_{ij}) | \lambda_1(t_{ij}) )} ,
\end{array}
$$
$\mathbb{E} \left[ {c}_1(t_{ij}) | \mathbf{y}_{i}, \mathbf{X}_{i}, {\Theta}_{i} \right] = \text{Pr} \left( A(t_{i(j-1)})=1, A(t_{ij}) = 1 | \mathbf{y}_{i}, \mathbf{X}_{i}, {\Theta}_{i} \right)$, $\Theta_{i} = \left\{ \boldsymbol{\delta}_{i}, \boldsymbol{\beta}_s, \mathbf{b}_s, \sigma^2_s, \mu_s \right\}$ and
$$ 
\begin{array}{rl}
{\alpha}_s (t_{ij}) &\propto \text{Pr} \left( A(t_{ij})=s | y_{i0}, \dots, y_{ij}, \mathbf{x}_{i1}, \dots, \mathbf{x}_{ij}, {\Theta}_{i} \right) \\ 
{\nu}_s (t_{ij}) &\propto \text{Pr} \left( A(t_{ij})=s | y_{ij}, \dots, y_{in_{i}}, \mathbf{x}_{i(j+1)}, \dots, \mathbf{x}_{in_{i}}, {\Theta}_{i} \right)
\end{array}
$$ are forward and backward probabilities of a HMM. Vectors $\boldsymbol{\delta}_{i} \in \mathbb{R}_{\geq0}^{1 \times 2}$ are of initial state distribution probabilities for the HMM. The transition probability matrix $\boldsymbol{\Gamma}(t_{ij}) \in \mathbb{R}_{\geq0}^{2 \times 2}$, derived by normalizing the alternating recurrent event exponential PH models are
$$
\boldsymbol{\Gamma}(t_{ij}) = 
\begin{bmatrix} \gamma_{11}(t_{ij}) & \gamma_{12}(t_{ij}) \\
\gamma_{21}(t_{ij}) & \gamma_{22}(t_{ij})
\end{bmatrix} =
\begin{bmatrix} 1-\text{expit}\left( \eta_1(t_{ij}) \right) & \text{expit}\left( \eta_1(t_{ijk}) \right) \\
\text{expit}\left( \eta_2(t_{ij}) \right) & 1 - \text{expit}\left( \eta_2(t_{ij}) \right)
\end{bmatrix}
$$
where $f ( \Delta(t_{ij}) | \lambda_s(t_{ij}) ) / \left\{ f ( \Delta(t_{ij}) | \lambda_s(t_{ij}) ) + S ( \Delta(t_{ij}) | \lambda_s(t_{ij}) ) \right\} = \left\{ 1 + \exp(-\eta_s(t_{ij})) \right\}^{-1} =\\ \text{expit}\left( \eta_s(t_{ij}) \right)$. The weights from $\mathbf{w}(t_{ij})$ can more generally be written as \\$\text{Pr}\left( A(t_{i(j-1)})=q, A(t_{ij})=r | \mathbf{y}_{i}, \mathbf{X}_{i}, {\Theta}_{i} \right) \propto  \alpha_q (t_{i(j-1)}) \nu_r (t_{ij}) \gamma_{qr} (t_{ij})$. The E-step involves imputing $\mathbf{w}(t_{ij})$ through a HMM, using a forward-backward algorithm, where transition probabilities are standard logistic functions. We denote forward and backward probabilities vectors
$\boldsymbol{\alpha}^\top (t_{ij}) = \boldsymbol{\delta}_{i} \mathbf{P}(t_{i0}) \prod^{j}_{m=1} \boldsymbol{\Gamma} (t_{im}) \mathbf{P}(t_{im})$ and $
\boldsymbol{\nu} (t_{ij}) = \mathbf{P}(t_{ij}) \prod^{n_{i}}_{m=j+1} \boldsymbol{\Gamma} (t_{im}) \mathbf{P}(t_{im}) \mathbf{1} $, where $\boldsymbol{\alpha} (t_{ij})$ and $\boldsymbol{\nu} (t_{ij})$ are $2 \times 1$ vectors. The state dependent distribution are contained in the $2 \times 2$ diagonal matrix $\mathbf{P}(t_{ij})= \text{diag} \Big( f \left(y(t_{ij}) | \mu_1 \right), f \left(y(t_{ij}) | \mu_2 \right) \Big)$. Note that $\text{Pr}(\mathbf{Y}_{i} =\mathbf{y}_{i} | \mathbf{X}_{i}, {\Theta}_{i}) = \boldsymbol{\alpha}^\top (t_{ij}) \boldsymbol{\Gamma} (t_{ij}) \boldsymbol{\nu} (t_{i(j+1)})$, and are used to normalize E-step probabilities. The E-step update for iteration $l+1$, simplifies to calculating probabilities $\mathbf{w}^{(l+1)}(t_{ij})$ as
$$
\begin{bmatrix}
c^{(l+1)}_1(t_{ij}) & d^{(l+1)}_1(t_{ij}) \\
d^{(l+1)}_2(t_{ij}) & c^{(l+1)}_2(t_{ij})
\end{bmatrix} \propto \left(
{\boldsymbol{\alpha}^\top}^{(l)}(t_{i(j-1)}) \otimes \boldsymbol{\nu}^{(l)}(t_{ij}) \right) \odot \boldsymbol{\Gamma}^{(l)}(t_{ij})
$$
such that the sum of all elements is equal to one, $\| \mathbf{w}^{(l+1)}(t_{ij}) \|_1 = 1$. Operations $\otimes$ and $\odot$ are Kronecker and Hadamard products respectively. The update for the Poisson mixture model weights are $\mathbf{u}^{(l+1)} (t_{ij}) \propto \boldsymbol{\alpha}^{(l)}(t_{ij}) \odot \left( \boldsymbol{\Gamma}^{(l)}(t_{i(j+1)}) \boldsymbol{\nu}^{(l)}(t_{i(j+1)}) \right)$, such that $\| \mathbf{u}^{(l+1)}(t_{ij}) \|_1 = 1$.

\subsubsection{M-step} 
\label{MStep}

The M-step update for $\left\{ \mu_1^{(l+1)}, \mu_2^{(l+1)} \right\}$ involves solving the mixture model
$L \left( \mu_1, \mu_2 | \mathbf{u}^{(l+1)} \right) = \prod^{n_{i}}_{j=0} \prod^{2}_{s=1} f \left( y(t_{ij}) | \mu_s \right)^{u^{(l+1)}_s(t_{ij})}$, where solutions are known for most distributions and $\mu^{(l+1)}_s = \left({ \sum_{ij} u_s^{(l+1)}(t_{ij}) } \right)^{-1} \left({ \sum_{ij} u_s^{(l+1)}(t_{ij}) y(t_{ij}) } \right)$ in the Poisson setting. From $\left\{ \mu_1^{(l+1)}, \mu_2^{(l+1)} \right\}$, we have updates $\mathbf{P}^{(l+1)}(t_{ij}) $. The update for $\left\{ \boldsymbol{\beta}_s^{(l+1)}, \mathbf{b}_s^{(l+1)}, {\sigma^2_s}^{(l+1)} \right\}$ involves fitting two frailty models for $L \left (\boldsymbol{\beta}_s, \sigma^2_s, \mathbf{b}_s | \mathbf{c}_s^{(l+1)}, \mathbf{d}_s^{(l+1)} \right)$ which can be accomplished by recognizing $\left\{ \mathbf{d}_s^{(l+1)}, \mathbf{c}_s^{(l+1)}\right\}$ can be factored into indicators and log-likelihood weights or case weights. The weights can interpreted as the probability that a specific state transition, e.g., if $\widehat{d}_1(t_{ij})\approx1$, then an active-to-rest transition likely occurred. We duplicate each row of data to be both a transition event and a censored outcome and then weight the rows by $\mathbf{d}_s^{(l+1)}$, and $\mathbf{c}_s^{(l+1)}$ respectively. A data augmentation example is outlined in Web Table 1.

While there are numerous survival packages in \texttt{R}, we need a package that can incorporate weights, parametric PH models and normally distributed random intercepts \citep{therneau2015package}.
The \texttt{R} package \texttt{tramME} can be used to fit our weighted exponential parametrization of shared log-normal frailty models \citep{tamasi2021tramme}. The package \texttt{tramME} uses a transformation model approach combined with an efficient implementation of the Laplace approximation to fit the shared log-normal frailty models \citep{hothorn2018most,hothorn2020most,kristensen2016tmb}. 
By updating $\left\{ \boldsymbol{\beta}_1^{(l+1)}, \mathbf{b}_1^{(l+1)}, \boldsymbol{\beta}_2^{(l+1)}, \mathbf{b}_2^{(l+1)} \right\}$ we also update our transition rates, $\lambda^{(l+1)}_1(t_{ij})$ and $\lambda^{(l+1)}_2(t_{ij})$ which are used to calculate $\boldsymbol{\Gamma}^{(l+1)} (t_{ij})$. 

The M-step updates for the initial state distribution are $\boldsymbol{\delta}^{(l+1)}_{i} \propto \\ \left( \boldsymbol{\delta}^{(l)}_{i} \mathbf{P}^{(l+1)}(t_{i0}) \right) \odot \left( \boldsymbol{\Gamma}^{(l+1)}(t_{i1}) \boldsymbol{\nu}^{(l+1)}(t_{i1}) \right) $, such that $\| \boldsymbol{\delta}^{(l+1)}_{i} \|_1 = 1$. Finally, we iteratively calculate the E-step: $\left\{ \mathbf{w}^{(l+1)}(t_{ij}), \mathbf{u}^{(l+1)}(t_{ij}) \right\}$, and M-step: $\left\{ \boldsymbol{\delta}^{(l+1)}_{i}, \boldsymbol{\beta}^{(l+1)}_s, \mathbf{b}^{(l+1)}_s, {\sigma^2}^{(l+1)}_s, \mu^{(l+1)}_s \right\}$ until convergence to obtain the maximum likelihood estimates. 
Estimating shared population parameters can be done by taking the product of all individual likelihoods 
and applying EM.

\subsection{Comparison of Relative Risks from PH and Logistic Regression}
\label{penalty}
The estimated relative risk or coefficient estimates from PH and logistic regression have been shown to be similar under a variety of situations \citep{abbott1985logistic,ingram1989empirical,thompson1977treatment,callas1998empirical}. Many useful properties of PH modeling can be leveraged in Markov chains which can improve the robustness of parameter estimation and reduce computational burden. Though many analyses look at the Cox PH case, we can adapt these approaches to the exponential PH which is a special case of the Cox PH. Following the derivations of \cite{abbott1985logistic}, we have $\lambda(t)=-d \log \{S(t)\} / d t$ and $S( \Delta(t) ) =\exp \left\{ -\exp \left(\zeta_0( \Delta(t) )+\sum_k \beta_{k} x_{k}\right) \right\}$ with $\zeta_0 (\Delta (t)) =\log \left\{ \int_0^{ \Delta(t) } \lambda_{0}(z) d z \right\} = \log(\Delta(t)) + \beta_0 $ where $\lambda_{0}(z) = \exp(\beta_0)$ is the baseline hazard from the exponential distribution. When the event times are discrete, $\Delta( t ) = 1$ can be arbitrarily assigned as the event time and $\zeta_0 ( \Delta(t) ) = \beta_0$. Under this setting, $\log \left\{ -\log \left( S(\Delta(t_{ij})) \right)\right\} = \eta_s(t_{ij})$ and the Taylor expansion of the survival function $S(\Delta(t_{ij})=1 \mid \lambda_s(t_{ij})) = \exp(-\lambda_s(t_{ij}))$ at $\lambda_s(t_{ij})=0$ results in the power series $
\exp(-\exp(\eta_s(t_{ij}))) 
= 1-\lambda_s(t_{ij})+\lambda_s(t_{ij})^{2} / 2 !-\lambda_s(t_{ij})^{3} / 3 ! \ldots+(-1)^{n} \lambda_s(t_{ij})^{n} / n ! \ldots 
= 1-\lambda_s(t_{ij}) + R_{\text{PH}} \left( \lambda_s(t_{ij}) \right)
$ where $1-P_{ij} = S(\Delta(t_{ij})=1 \mid \lambda_s(t_{ij})) \approx 1-\lambda_s(t_{ij})$ when the incidence rate $\lambda_s(t_{ij})\approx 0$. The respective power series of $1-\text{expit}(\eta_s(t_{ij}))$ is given as 
$
1-\text{expit}(\eta_s(t_{ij})) = 1-\lambda_s(t_{ij})+\lambda_s(t_{ij})^{2}-\lambda_s(t_{ij})^{3} \ldots+(-1)^{n} \lambda_s(t_{ij})^{n} \ldots 
= 1-\lambda_s(t_{ij}) + R_{\text{log}} \left( \lambda_s(t_{ij}) \right)
$ where we set $\exp(-\exp(\eta_s(t_{ij}))) = 1-\text{expit}(\eta_s(t_{ij})) = 1-P_{ij} \approx 1-\lambda_s(t_{ij})$ when the incidence rates are low. Note that in the case of low incidence rates, $R_{\text{PH}} \left( \lambda_s(t_{ij}) \right) \leq R_{\text{log}} \left( \lambda_s(t_{ij}) \right)$ and writing $\eta_s(t_{ij})$ as a function of $R_{\text{PH}} \left( \lambda_s(t_{ij}) \right)$ or $R_{\text{log}} \left( \lambda_s(t_{ij}) \right)$, we get $\log \left( P_{ij} + R_{\text{PH}} \left( \lambda_s(t_{ij}) \right) \right) \leq \log \left( P_{ij} + R_{\text{log}} \left( \lambda_s(t_{ij}) \right) \right)$. As noted in previous studies, PH and logistic regression estimate similar relative risk under a group event time, low incidence rate setting \citep{abbott1985logistic,ingram1989empirical}. In many practical settings, discrete-time applications of binary outcome models involve low incidence rates, letting exponential PH regression serve as an alternative to logistic regression. The analogous Markov chain with rare outcomes is a process with an extended stay in the current state, commonly found in diurnal biological processes. However, when event times are heterogeneous, then PH regression is the more appropriate choice for estimating relative risk. As noted in \cite{abbott1985logistic}, in many cases, we observed shrinkage towards zero for the relative risk under PH regression when compared to the logistic regression due to the inequality between the remainder terms $R_{\text{PH}} \left( \lambda_s(t_{ij}) \right)$ and $R_{\text{log}} \left( \lambda_s(t_{ij}) \right)$ \citep{ingram1989empirical,thompson1977treatment,callas1998empirical}.

We further expand on this observed shrinkage by showing that the exponential PH regression is a penalized logistic regression. The exponential canonical form of exponential PH and logistic regression is given as $
L\left( \boldsymbol{\beta}_s \mid \mathbf{A}_i \right) = \prod^{n_{i}}_{j=1} \exp \left\{ d_s(t_{ij}) \eta_s(t_{ij}) \right\} \exp \left\{ -\Psi\left( \eta_s(t_{ij}) \right) \right\}, \ 
$ with $d_s(t_{ij})\in\{0,1\}$. Here, $\Psi_{\text{log}}(\eta) = \log(1+\exp(\eta))$ for logistic regression and $\Psi_{\text{PH}}(\eta) = \Delta (t) \exp(\eta) = \exp(\eta)$ for exponential PH regression under the discrete-time setting. We see that the logistic regression likelihood is uniformly bounded below by the PH likelihood due to $\exp(-\log(1+\exp(\eta))) >  \exp(-\exp(\eta))$, which simply follows from the inequality $\log(1+z) < z$ for $z>0$. The convex optimization problems of PH and logistic regression are to minimize loss functions: $J_{\text{log}}(\boldsymbol{\beta}_s ) = - \log L_{\text{log}} \left( \boldsymbol{\beta}_s \mid \mathbf{A}_i \right) $ and $J_{\text{PH}}(\boldsymbol{\beta}_s) = - \log L_{\text{PH}} \left( \boldsymbol{\beta}_s \mid \mathbf{A}_i \right)$. There is a positive convex penalty difference between the logistic and PH optimization problems, $
J_{\text{PH}}(\boldsymbol{\beta}_s) = J_{\text{log}}(\boldsymbol{\beta}_s) + \mathcal{P}(\boldsymbol{\beta}_s) $ and $
\mathcal{P}(\boldsymbol{\beta}_s) = \sum^{n_{i}}_{j=1} -\log(1+ \exp( \eta_s(t_{ij}) ) ) + \exp( \eta_s(t_{ij}) ) > 0
$ 
with the penalty as the difference of $\Psi_{\text{log}}(\eta)$ and $\Psi_{\text{PH}}(\eta)$. It is straight forward to see that $ \mathcal{P}(\boldsymbol{\beta}_s)$ is convex with hessian $\nabla_{\boldsymbol{\beta}_s}^2 \mathcal{P}(\boldsymbol{\beta}_s) = \mathbf{X}^\top \boldsymbol{\Omega} \mathbf{X} \succeq 0$ and $\boldsymbol{\Omega}$ as a diagonal matrix with elements $\exp(\eta_s(t_{ij})) \left[ 1 - \{1+ \exp(\eta_s(t_{ij}))\}^{-2} \right] > 0$. Under a simple parameterization, such as the intercept only model, it follows that the penalty $\mathcal{P}(\boldsymbol{\beta})$ favors a low incidence rate. The function $-\log(1+ \exp(\eta)) + \exp(\eta)$ is relatively flat for $\eta<0$ and dominated by $\exp(\eta)$ for $\eta \gg 0$. However, in practice this will shrink coefficients to the solution of $\text{argmin}_{\boldsymbol{\beta}_s} \{ \mathcal{P}(\boldsymbol{\beta}_s) \}$ which tends to be near $\boldsymbol{\beta}_s=0$ when considering that each $\eta_s(t_{ij})= \mathbf{x}^\top (t_{ij}) \boldsymbol{\beta}_s$ is a different linear combination of $\boldsymbol{\beta}_s$. The penalty $\mathcal{P}(\boldsymbol{\beta}_s)$ modifies the convex hull of the logistic regression loss function to induce shrinkage and results in a PH loss function which has many readily available software for estimation.

As noted in our E-step, the conditional expectations of the EM algorithms reduces to transition probability matrices that are comprised of logistic regression probabilities. In the case of a HMM with 3 or more states, the E-step reduces to transition probabilities comprised of multinomial logistic regression probabilities. However, the M-step conveniently involves fitting several independent exponential PH models. In order to illustrate this point, we define the complete data likelihood of transitioning out of state 1 in a 3 state HMM as, 
\begin{equation} \label{HMM3states}
\begin{array}{rl}
L( \boldsymbol{\beta}_{12}, \boldsymbol{\beta}_{13} | \mathbf{A}_i ) &= L( \boldsymbol{\beta}_{12} | \mathbf{A}_i ) L( \boldsymbol{\beta}_{13} | \mathbf{A}_i )  \\
&= \prod^{n_{i}}_{j=1}
\left\{ \lambda_{12}(t_{ij}) \right\}^{d_{12}(t_{ij})}
\left\{ \lambda_{13}(t_{ij}) \right\}^{d_{13}(t_{ij})}
S ( \Delta(t_{ij}) | \sum_{k=2}^3 \lambda_{1k}(t_{ij}) )
\end{array}
\end{equation}
with $d_{1m}(t_{ij}) = \mathbb{I} \left[ A(t_{i(j-1)})=1, A(t_{ij}) = m \right]$ for $m\in \{1,2,3\}$, $\sum_{m=1}^3 d_{1m}(t_{ij}) = 1$ and $\lambda_{1k}(t_{ij}) = \exp( \eta_{1k}(t_{ij}) ) = \exp( \mathbf{x}^\top (t_{ij}) \boldsymbol{\beta}_{1k} )$ for $k \in \{ 2, 3 \}$. The minimum of two or more independent exponential random variables follows an exponential distribution with a new rate equal to the sum of rates, in our case: $\sum_{k=2}^3 \lambda_{1k}(t_{ij})$. The likelihood contribution of staying in the same state is the survival function $S \left( \Delta(t_{ij}) | \sum_{k=2}^3 \lambda_{1k}(t_{ij}) \right) = \prod_{k=2}^3 S \left( \Delta(t_{ij}) | \lambda_{1k}(t_{ij}) \right) = \prod_{k=2}^3 \exp( \Delta(t_{ij}) \lambda_{1k}(t_{ij}) )$. Continuing our example, calculating the E-step  for the transition from state 1 to 2 is given as multinomial probability $\mathbb{E} \left[ d_{12}(t_{ij}) \right] = \lambda_{12}(t_{ij}) / \left\{ 1 + \sum_{k=2}^3 \lambda_{1k}(t_{ij}) \right\}$ where we leave the derivation details to the Web Appendix B. Note that in the 3 state HMM, E-step imputations are constrained to a 9-dimensional simplex. The M-step for the 3 state HMM parameters from likelihood \eqref{HMM3states}, can be estimated with 2 independent exponential PH models. This EM procedure can be extended to an arbitrary number of states. In the M-step we fit independent weighted PH models rather than a cumbersome multinomial regression.

As noted in previous work, when the event times are heterogenous, PH regression is the correct model for estimating relative risk \citep{abbott1985logistic,ingram1989empirical,thompson1977treatment,callas1998empirical}. However, PH and logistic regression estimate similar relative risk for the discrete and grouped event time setting with low incidence rates. We showed that under the discrete-time setting, exponential PH regression is an implicitly penalized logistic regression resulting in shrinkage of the relative risk estimates. This is desirable in many situations, specifically in our case where we are building HMMs, a model with a complicated error in response mechanism. The penalty $\mathcal{P}(\boldsymbol{\beta}_s)$ slightly favors low probabilities of transitioning out of a state, which is useful to mitigate false positives. As a result, penalization shrinks the transition probability matrices $\boldsymbol{\Gamma} (t_{ij})$, towards the identity matrix and favors an extended sojourn time. Of note, post estimation HMM procedures such as the Viterbi algorithm for finding the most likely path also favors low probabilities of transitioning out of a state \citep{viterbi1967error,forney1973viterbi}. In addition, we also invoke mixed effect modeling which numerically benefits from penalization. Next, fitting multiple independent weighted frailty models in the M-step is a more straight forward procedure than fitting an analogous weighted mixed multinomial logistic regression. The PH model is the correct model for heterogenous event times, but even in the case of discrete-time transitions, PH modeling's implicit penalization has several useful properties for estimating HMMs.

\subsection{Competing Methods}
\label{comp}
We denote the alternating recurrent event exponential PH model outlined in Section \ref{est} as PH-HMM. For our competing methods, we use the CT-HMM and discrete-time mixed effect logistic regression HMM (denoted as DT-HMM), described in our Web Appendix A. In addition, we define a two step estimator based on Poisson mixture model (PMM) \textit{maximum a posteriori} (MAP) estimates in the Web Appendix A. All competing methods are initialized using PMM MAP estimates for state labels and EM is repeated until $\| \Theta^{(l+1)} - \Theta^{(l)} \|_1 \leq 10^{-4}$. 

\section{Results}
\label{results}

The EM algorithm converges and obtains the MLEs $\left\{ \widehat{\boldsymbol{\beta}}_s, \widehat{\mathbf{b}}_s, \widehat{\sigma}^2_s, \widehat{\mathbf{c}}_s, \widehat{\mathbf{d}}_s \right\}$, with the final likelihood evaluation being equivalent to weighted exponential frailty models. As a result, we can generalize existing large sample theory asymptotic inference for mixed effect models for our weighted setting, using weights derived in Section \ref{EStep}. 

\begin{theorem} \label{thm1}
Regression coefficients $\boldsymbol{\beta}_s$ are asymptotically normally distributed 
$\widehat{\boldsymbol{\beta}}_s \stackrel{D}{\sim}  \mathrm{N} \left( \boldsymbol{\beta}_s, \Sigma_s \right)$ 
where $\Sigma_s$ are corresponding $\boldsymbol{\beta}_s$ elements of the inverse observed informations $\mathcal{I}^{-1}(\boldsymbol{\beta}_s, \mathbf{b}_s) $. The observed informations are given as $
\mathcal{I}(\boldsymbol{\beta}_s, \mathbf{b}_s) 
= \\
\mathbf{U}^\top
\left(
- \nabla^2_{\boldsymbol{\eta}_s} \log L(\boldsymbol{\beta}_s | \mathbf{b}_s, \widehat{\mathbf{c}}_s, \widehat{\mathbf{d}}_s )
\right)
\mathbf{U} + \mathrm{diag}(\mathbf{0}_{ p \times p}, \sigma_s^{-2} \mathbf{I} ) $
where  
$
- \nabla^2_{\boldsymbol{\eta}_s} \log L(\boldsymbol{\beta}_s | \mathbf{b}_s, \widehat{\mathbf{c}}_s, \widehat{\mathbf{d}}_s ) 
= \\
\textup{diag} \big( 
\Delta(t_{i1}) \lambda_s (t_{i1}), \Delta(t_{i1}) \lambda_s (t_{i1}), \dots,  \Delta(t_{ij}) \lambda_s(t_{ij}), \Delta(t_{ij}) \lambda_s(t_{ij}), \dots
\big) \mathbf{W}_s $, 
$$
{\mathbf{U}^\top 
=
\left[ \begin{array}{llllllll}
\mathbf{x} (t_{i1}) & \mathbf{x} (t_{i1}) & \mathbf{x} (t_{i2}) & \mathbf{x} (t_{i2}) & \dots &
\mathbf{x} (t_{ij}) & \mathbf{x} (t_{ij}) & \dots \\
\mathbf{z} (t_{i1}) & \mathbf{z} (t_{i1}) & \mathbf{z} (t_{i2}) & \mathbf{z} (t_{i2}) & \dots &
\mathbf{z} (t_{ij}) & \mathbf{z} (t_{ij}) & \dots 
\end{array} \right]} 
$$ 
and $
\mathbf{W}_s =
\textup{diag} \big(
\widehat{d}_s(t_{i1}), \widehat{c}_s(t_{i1}), \widehat{d}_s(t_{i2}), \widehat{c}_s(t_{i2}),\dots, \widehat{d}_s(t_{ij}), \widehat{c}_s(t_{ij}),\dots 
\big)$.
\end{theorem}

\subsection{Simulation Studies}
\label{sim_study}
For the alternating event process, we use a 24h period sine function as our time varying covariate, $x(t) = \sin \left( {2 \pi t}/{24} \right) \in [-1,1]$ and independently draw censoring times from a uniform distribution, $r_{ij} \sim U(0,h_{\text{max}})$. For the PH models we sequentially increment $t_{ij}$ by $\Delta(t_{ij})$ and simulate covariates $x(t_{ij})$. We simulate the complete data likelihood with multiple individuals but shared $\boldsymbol{\beta}_s$, by drawing from
$$
\begin{array}{l}
\lambda_1(t_{ij}) \mid ( A(t_{i(j-1)}) = 1 ) = \exp \left( \eta_1(t_{ij}) \right) =  \exp \left( \mathbf{x}^\top (t_{ij}) \boldsymbol{\beta}_1\right) = \exp \left( \beta_{10} + \beta_{11} x(t_{ij}) \right) 
\end{array}
$$ and 
$\lambda_2(t_{ij}) \mid ( A(t_{i(j-1)}) = 2 ) = \exp \left( \beta_{20} + \beta_{21} x(t_{ij}) \right) $ with $v_s(t_{ij}) \sim \text{Exp} \left( \lambda_s(t_{ij}) \right)$, $\Delta(t_{ij}) = \min \{v_s(t_{ij}), r_{ij}\}$, and $d_s (t_{ij}) = \mathbb{I}[ \Delta(t_{ij}) = v_s(t_{ij}) ] = \mathbb{I}[ A(t_{i(j-1)})=s, A(t_{ij}) \ne s ]$. Pseudocode for generating the alternating survival data can be found in Algorithm 1 of the Web Appendix. Similarly, for a discrete-time alternating event process, we simulate transition events as $d_s (t_{ij}) \sim \text{Bernoulli}\left( \text{expit}(\eta_s(t_{ij})) \right)$ and increment $t_{ij}$ by $\Delta(t_{ij})=1$. Pseudocode for generating the discrete alternating event data can be found in Algorithm 2 of the Web Appendix. Once true state labels $A(t_{ij})$ are obtained, we simulate state dependent observations from Poisson distributions to complete the simulation of the HMMs.

We simulated 500 replicates using four different sets of parameters and used 3 methods for generating the data. For each set of parameters, we simulated data using Algorithm 1 with maximum censoring times set to $h_{\text{max}}=10$ and $h_{\text{max}}=1$ to evaluate performance under heterogeneous and grouped event times. We also used Algorithm 2 in order study models fitted to data simulated from a discrete-time process. We have a total of 12 different cases outlined in Table \ref{sim}. In summary, Cases 1.1--1.3 looked at low incidence rate of state transition and a large distributional difference between $f (y(t_{ij}) | {\mu_1})$ and $f (y(t_{ij}) |{\mu_2})$; Cases 2.1--2.3 looked at high incidence rate and large distributional difference; Cases 3.1--3.3 looked at low incidence rate and small distributional difference; and Cases 4.1--4.3 looked at high incidence rate and small distributional difference. For each case we fit models using the PMM, DT-HMM, CT-HMM and PH-HMM approaches. The E-step update can be computed quickly in parallel where each $i$th HMM is processed independently. We present mean accuracy of recovering the true state labels using the MAP and empirical standard error (SE), over 500 replicates in Table \ref{sim}. We also present mean parameter estimates, empirical standard errors and mean square errors (MSE) in Tables \ref{sim_beta} and \ref{sim_etc}.

First, we present our findings regarding the accuracy of recovering the true label. When the difference between state dependent distributions is large, such as Cases 1.1--1.3 and 2.1--2.3, all competing methods have comparable accuracy. In Cases 3.1--3.3 and 4.1--4.3, with a small distributional difference, we observe lower accuracy in PMM. In Cases 3.2 and 4.2, with low incidence rates and grouped event times, DT-HMM, CT-HMM and PH-HMM have comparable accuracy for reasons outlined in Section \ref{penalty}. When $\beta_{11}$ and $\beta_{21}$ have opposite signs, the range of $\Delta(t_{ij})$ is small and there is a low incidence, then the estimated CT-HMM transition probabilities from the matrix exponential, $\boldsymbol{\Gamma} (t_{ij}) = \exp \left( \mathbf{Q}(t_{ij}) \Delta(t_{ij}) \right)$, approximates the transition probability matrices from DT-HMM and PH-HMM. As a results, in these cases, DT-HMM, CT-HMM and PH-HMM have comparable accuracy. However, given heterogeneous event times and high incidence rates (Case 4.1), the model misspecification of CT-HMM noticeably reduces accuracy and is also reflected in the high MSE of parameter estimates. In the case of small distributional difference and heterogeneous event times (Cases 3.1 and 4.1), we observed that DT-HMM was more accurate than PH-HMM but had greater bias in its coefficients estimates (see Tables \ref{sim_beta} and \ref{sim_etc}).

We outline our findings regarding heterogeneous event times, $\Delta(t_{ij}) \in [0,h_{\text{max}}]$ for $h_{\text{max}}=10$ (Cases 1.1, 2.1, 3.1 and 4.1). In these cases, we observed that DT-HMM has poor performance in estimating the relative risk, specifically the baseline risk or intercept. In Cases 1.1, 2.1, 3.1 and 4.1 we observed large MSE in the DT-HMM coefficient estimates, emphasizing discrete-time based approaches are sensitive to heterogeneous event times. Of note, in Case 2.1: heterogeneous event times, high incidence rates and large distributional difference, PH-HMM clearly out performs CT-HMM and DT-HMM. This finding is inline with Section \ref{penalty}, where we noted that DT-HMM and PH-HMM results are similar in grouped event time low incidence setting. 

Next, we discuss additional implications of the derivations from Section \ref{penalty} in our simulation study. When evaluating parameter estimation, we observed higher MSE of $\mu_1$ and $\mu_2$ for PMM in a small distributional difference settings (Cases 3.1--3.3 and 4.1--4.3). The PMM method is a two step estimator and does not use covariate information to estimate $\mu_1$ and $\mu_2$, leading to highly variable estimates. Because PMM does not use $\{ \mu_1, \mu_2 \}$ and $\boldsymbol{\beta}_s$ simultaneously during estimation, PMM estimates are generally less accurate than other competing methods. In the cases with low incidence rate, discrete and grouped event times (Cases 1.2, 1.3, 3.2 and 3.3), we observed that DT-HMM, CT-HMM and PH-HMM yielded similar parameter estimates. In cases with a small distributional difference, we observed that all estimates become less accurate when compare with a large distributional difference setting. When considering a grouped event time, low incidence rate with PH data generating process (Cases 1.2 and 3.2) we observed that the relative risk from DT-HMM are inflated when compared the truth and PH-HMM estimates. We may contrast these results with the cases where the data is generated from a logistic regression (Cases 1.3, 2.3, 3.3 and 4.3). DT-HMM is best suited for estimation when the underlying process is a logistic regression. However, in Cases 1.3, 2.3, 3.3 and 4.3, PH-HMM yields estimates where $\beta_{10}$ and $\beta_{20}$ are pulled towards $-\infty$ and $\beta_{11}$ and $\beta_{21}$ are pulled towards 0. These findings are aligned with the derivations found in Section \ref{penalty}. This shrinkages slightly favors lower probabilities of transitioning out of a state, i.e., shrinking state transition matrices towards the identity matrix. In other words, PH models penalizes HMMs generated from a discrete-time processes to favor an extended sojourn time while maintaining comparable estimates for $\mu_s$. This shrinkage is useful for mitigating overfitting, especially when the goal is to fit a complicated HMM with no prior knowledge on the true data generating process.

From our simulation studies, we found that PH-HMM excels in a number of different cases. DT-HMM and CT-HMM are sensitive to heterogeneous event times leading to inaccurate state label recovery and bias parameter estimation. While DT-HMM is more accurate than PH-HMM at state label recovery in Cases 3.1 and 4.1, their parameter estimates have higher MSE and bias. DT-HMM generally has higher MSE than PH-HMM for cases where the data is generate from PH models (Cases 1.1, 1.2, 2.1, 2.2, 3.1, 3.2, 4.1 and 4.2). The shrinkage properties outlined in Section \ref{penalty} carries over into our simulation study. In most cases and when the data is generated from a logistic regression, DT-HMM and PH-HMM have comparable accuracy and state dependent distribution estimates, even though PH-HMM coefficients exhibit shrinkage. This shrinkage property has many uses which extend beyond simple simulation studies which we outline next in our real data analysis examples.

\subsection{Application: mHealth Data}
A sample of $I = 41$ individuals, recruited via the Penn/CHOP Lifespan Brain Institute or through the Outpatient Psychiatry Clinic at the University of Pennsylvania as part of a study of affective instability in youth \citep{xia2022mobile}. All participants provided informed consent to all study procedures. This study was approved by the University of Pennsylvania Institutional Review Board. For each individual, roughly 3 months of data was collected using the Beiwe platform \citep{torous2016new}. Accelerometer measures (meters per second squared) for x, y and z axes, screen-on events for Android devices and screen-unlock events for iOS (Apple) devices were acquired through the Beiwe platform–we refer to both as “screen-on events” in this manuscript. 
In general, we recommend a minimum of 30 days of data in order to fit our individual-specific model with hour-of-day random intercepts.

First, we analyzed the data under a discrete-time setting where we impute data for each missing hour. By collecting both screen-on events as well as accelerometer data, we are able to construct a missing at random (MAR) model for imputing missing accelerometer data. In our case, there is missing accelerometer data during of periods screen-on activity i.e., accelerometer data is missing over a given hour while there are many screen-on events observed over the same period. On the other hand, there are periods of dormancy: accelerometer data is missing and there are also no screen-on events. More specifically periods of dormancy occur when accelerometer measurements are missing due to user and device related factors such as the phone being powered off or being in airplane mode. Periods of dormancy have greater probability of missing accelerometer features and are identified using a two state hidden semi-Markov model with Bernoulli state dependent distributions \citep{bulla2010hsmm}. Missing mean acceleration magnitudes from dormant periods were imputed using the minimum of acceleration features (excluding outliers). While missing data assigned to the periods screen-on activity were imputed by regressing accelerometer features on $Y(t_{ij})$ over all hours where data is completely observed. For the heterogeneous event time example, we did not impute missing data and absorbed the duration of dormancy into the event times $\Delta(t_{ij})$. Periods of consecutive missing acceleration magnitudes over 24h constitutes an end of a Markov chain and the start of a new chain, where the likelihoods of multiple HMM sequences can be multiplied together for parameter estimation.

\subsubsection{Estimating Strength of Routine in Youth with Affective Disorders}
\begin{figure}
\centering
\includegraphics[width=6in]{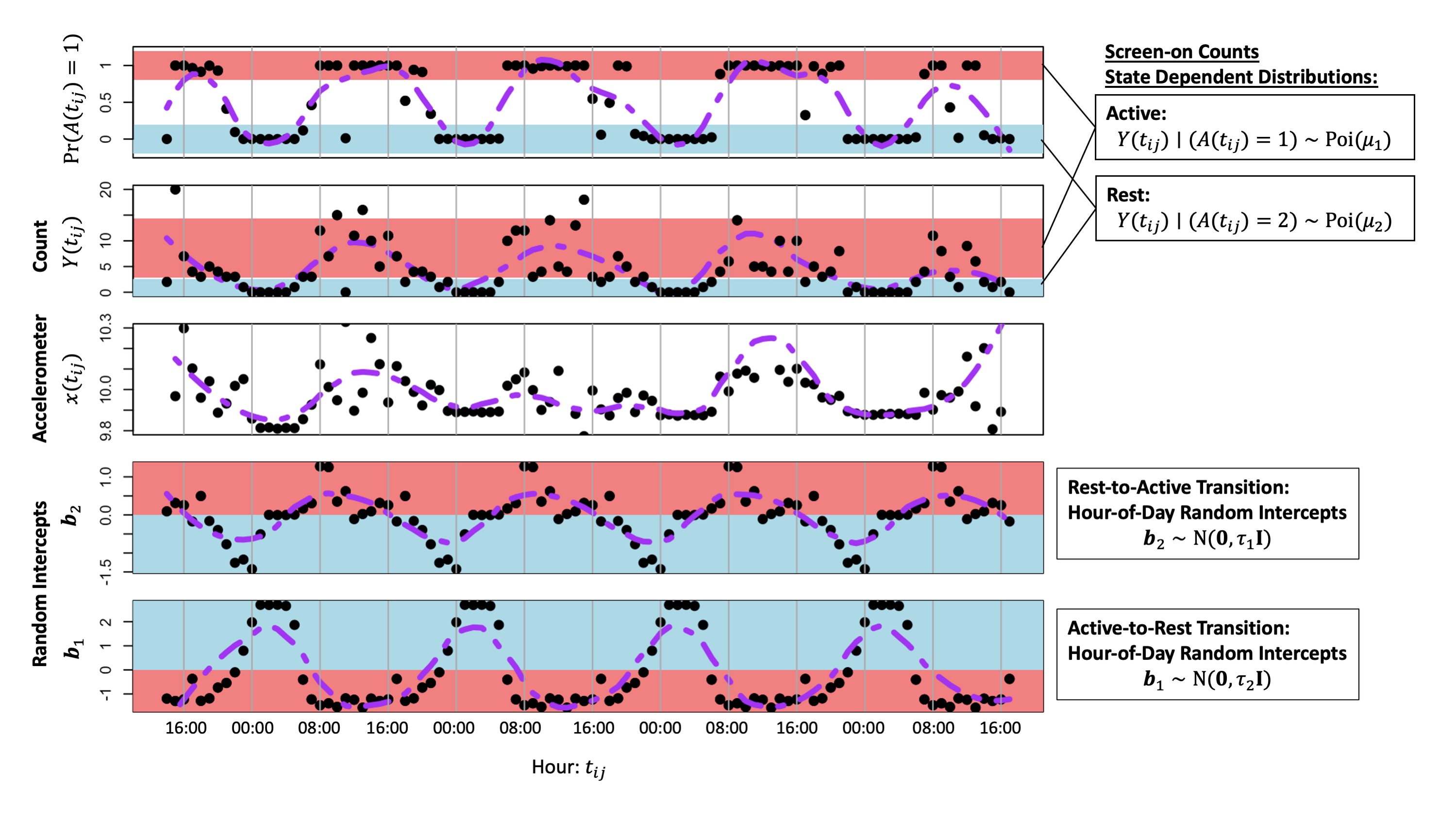}
\caption{{Example of Discrete-Time mHealth Data and Fitted PH-HMM}. Probabilities of being in the active state, screen-on counts, mean acceleration magnitude, and hour-of-day random intercepts are plotted against time (hours). Regression models: $\eta_s(t_{ij}) = \beta_{s0} + \beta_{s1} x^\top(t_{ij}) + \mathbf{z}^\top (t_{ij}) \mathbf{b}_{s}$, $\mathbf{b}_{s} \sim \mathrm{N}(\mathbf{0}_{24}, \tau_s \mathbf{I}_{24})$ were fitted for individual $i$ and MAP estimates were calculated using final E-step probabilities $\text{Pr}(A(t_{ij})=1)=\widehat{u}_1(t_{ij})$. Random intercepts capture the diurnal rhythm of active rest cycles, with active states mainly occurring between the hours of 6am-10pm. Large values of $\tau_s$ correspond with a high magnitude in the cyclic diurnal effects. }
\label{fig:3}
\end{figure}

In psychiatric studies, regularity of a rhythm is defined as the association of time-of-day and state membership; the effect of hour-of-day on activity and rest state membership represents the diurnal rhythm. As an illustrative example, we fit a model with hour-of-day effects, as normally distributed random intercepts in our HMMs and for each individual we fit PH models $\lambda_s(t_{ij}) = \exp \left( \mathbf{x}^\top(t_{ij}) \boldsymbol{\beta}_{s} + \mathbf{z}^\top (t_{ij}) \mathbf{b}_{s} \right)$, where $ \mathbf{b}_s \sim \mathrm{N} ( \mathbf{0}_{24}, \tau_s \mathbf{I}_{24} )$ are 24 hour-of-day random intercepts. By fitting hour-of-day random intercepts, $\mathbf{b}_1$ and $\mathbf{b}_2$ are each of length $24$ and $[\mathbf{z}(t_{ij})]_r=1$ only if $t_{ij}$ is in the $r$th hour of the day. Rates of transition from active-to-rest states are given as $\lambda_1(t_{ij})$, rates of transition from rest-to-active states are given as $\lambda_2(t_{ij})$ and an example of PH-HMM outputs can be found in Figure \ref{fig:3}. The variances of the random intercepts can be interpreted as a L2 penalty on hour-of-day effects and disappears as $\tau_s \rightarrow \infty$. We quantify the strength of diurnal effects for an individual, by looking at the variances $\tau_s$, where large variances correspond with large hour-of-day effect sizes and greater regularity in diurnal rhythms with an example in Figure \ref{fig:3}.

\subsubsection{Population HMM: Differences Between Operating Systems}
In addition, we can fit a population model, with random intercepts being specific to each individuals through estimation using the likelihood $\prod_{i=1}^I L( \boldsymbol{\beta}_1, \mathbf{b}_1, \sigma^2_1, \boldsymbol{\beta}_2, \mathbf{b}_2, \sigma^2_2, \mu_1, \mu_2 | \mathbf{A}_i )$. However, for iOS devices we only have screen-unlock events, i.e., entering in a passcode to unlock the phone. Android devices have screen-on events which occur when the phone screen turns on such as when receiving a message; the phone does not need to be unlocked for the screen to be turned on. Screen-unlock events are less frequent and a subset of screen-on events, causing the counts $Y(t_{ij})$, to be lower for iOS devices. The relationship between acceleration $x(t_{ij})$ and $Y(t_{ij})$, may be experience effect modification due to operating system (OS). We can test interaction between OS and acceleration, while controlling for the interaction with user sex in our regression and other individual effects with random intercepts. Android devices and males serve as baseline in this analysis. For the active-to-rest model: $\lambda_1(t_{ij})$ and rest-to-active model: $\lambda_2(t_{ij})$,
$
\lambda_s(t_{ij}) = \exp \left( \mathbf{x}^\top(t_{ij}) \boldsymbol{\beta}_{s} + \mathbf{z}^\top (t_{ij}) \mathbf{b}_{s} \right) = \exp \left( \beta_{s0} + \beta_{s1} x(t_{ij}) + 
\beta_{s2}x(t_{ij}) \mathrm{sex} + \beta_{s3}x(t_{ij}) \mathrm{OS} + 
\mathbf{z}^\top (t_{ij}) \mathbf{b}_{s} \right)$, where $\mathbf{b}_s \sim \mathrm{N} (\mathbf{0}_{41},\sigma^2_s \mathbf{I}_{41})$, $\mathbf{b}_{s}$ are 41 individual specific random intercepts and $\mathbf{z}^\top (t_{ij})$ are individual specific indicators. We fit our competing methods and test interaction for the discrete and heterogeneous event time settings, where estimates can be found in Table \ref{HMM_fit} and \ref{HMM_fit2}. 

We found that there was no significant interaction between OS and acceleration in the discrete-time active-to-rest model but did find significant interaction in the heterogeneous event time active-to-rest model. In addition, we found that there was significant interaction between OS and acceleration in the discrete-time rest-to-active model but did not find significant interaction in the heterogeneous event time rest-to-active model. During the active state, we know that counts $Y(t_{ij})$, are lower for iOS devices. In order to compensate for lower screen-on counts in iOS devices, iOS rest-to-active transitions require a higher magnitude of acceleration to achieve the same transition rate as an Android device, hence the negative sign of $\beta_{23}$. During the rest state, we know iOS devices have zero-inflated counts, where excess zeros are due to not being able to record a screen-on event. In the active-to-rest model, we may achieve the same transition rate while having higher acceleration in iOS devices than Android devices, hence the positive sign of $\beta_{13}$. Our results suggest that the magnitude of effect acceleration has on state transition depends on OS, but further investigation is needed. 

For the discrete-time setting, we estimated active state distribution $\mathbb{E}[y(t_{ij}) | A(t_{ij})=1] = \widehat{\mu}_1 \approx 8$ and rest state distribution $\mathbb{E}[y(t_{ij}) | A(t_{ij})=2] = \widehat{\mu}_2 \approx 0.4$. For the heterogeneous event time setting, we estimated $\mathbb{E}[y(t_{ij}) | A(t_{ij})=1] = \widehat{\mu}_1 \approx 9$ and $\mathbb{E}[y(t_{ij}) | A(t_{ij})=2] = \widehat{\mu}_2 \approx 1$. Screen-on counts separate into stark clusters where $\mu_1$ and $\mu_2$ are similar to the large distributional difference from our simulation study. We found that rate of transition from active-to-rest states are negatively associated with acceleration; rate of transition for rest-to-active states are positively associated with acceleration and are statistically significant ($p<0.05$) for all competing methods. These HMM parameter estimates related to screen-on counts and acceleration align with common intuition. For the PMM method, modeling the MAP estimates first and then combining the estimates to obtain a population level model, does not account for acceleration when imputing state labels and resulted in a poor fit. CT-HMM and PH-HMM estimates are comparable and aligned with the findings from the simulation study. In addition to large distributional differences, diurnal active-rest cycles are expected to be low incidence rate processes, where we anticipate a few transition between states in a 24 hour period. We observed that the magnitude of the relative risk estimates for CT-HMM and PH-HMM are comparable to each other but less than the DT-HMM estimates. This difference becomes more noticeable for the heterogeneous event time setting, which DT-HMM is not equipped to handle. For many parameters, DT-HMM relative risk estimates are 3 times that of PH-HMM estimates, while estimates for state dependent distribution parameters $\{ \mu_1, \mu_2 \}$ are similar. PH-HMM allows us to achieve comparable estimation of state dependent distributions while leveraging shrinkage to avoid overfitting coefficients. 

\vspace*{-1cm}
\section{Discussion}
\label{discuss}

For a latent state setting, we proposed a method for estimating alternating recurrent event exponential PH model with shared log-normal frailties using the EM algorithm. Our E-step imputations involves a discrete-time HMM using logistic or multinomial regression transition probabilities with normally distributed random intercepts. The HMM obtained during the E-step of our EM algorithm is an alternative method for estimating mixed hidden Markov models which are typically obtained by estimating logistic or multinomial regressions \citep{altman2007mixed,maruotti2012mixed}. The M-step conveniently involves fitting several independent PH models rather than multinomial regression with many states. In addition, we showed that the PH model applied to the discrete-time setting is a penalized logistic regression which shrinks transition probability matrices towards the identity matrix. Our framework can accommodate random intercepts to account for longitudinal data, such as data collected from the same individual or hour-of-day periodic effects. We derived asymptotic distributions for the PH regression coefficients and random intercepts, where coefficients have a hazard ratio interpretation akin to the Cox PH model. Our PH-HMM approach is a flexible method for modeling complex mHealth datasets, where heterogeneous event times can be incorporated into the PH regression while accounting for latent states. If the underlying data is a discrete-time process, then PH modeling offers slight penalization to mitigate overfitting, otherwise PH models are more appropriate for heterogeneous event time processes. 

We estimated two models in our real data analysis: one with missing data being accounted for through heterogeneous event times and another where missing data was imputed. By taking advantage of the fact that screen activity and accelerometer data are highly associated as well as the fact that screen activity data is never missing, the MAR assumption in our imputation model is well founded. That being said, if there's any residual explanation in the missing data even after accounting for the screen activity an MNAR and its corresponding sensitivity analyses would be more appropriate. We presented a flexible regression procedure that can accommodate different parameterization of random effects and the use of other statistical structures such as semiparametric regression and multilevel models. However, computational complexity and interpretability should be considered when parameterizing complicated models such as HMMs. A key advantage of our method is that complicated statistical structures can be incorporated into independent PH regressions which then simplifies to multinomial transition probabilities during the E-step. Though model selection statistics such as AIC/BIC was not explored in our manuscript, we evaluated the practical implications of PH modeling with a variety of different criteria. Simulation results and mHealth data analysis suggest that our PH-HMM excels in a variety of situations.


\backmatter


\section*{Acknowledgements}
This work was supported by grants from the NIMH: R01MH116884. Support for data collection was provided by the AE Foundation, R01MH113550 \& R01MH107703. The authors wish to thank the Lifespan Informatics \& Neuroimaging Center for providing the data


%

\bibliographystyle{biom} 
\bibliography{references.bib}

\begin{thebibliography}{}

\bibitem[\protect\citeauthoryear{Abbott}{Abbott}{1985}]{abbott1985logistic}
Abbott, R.~D. (1985).
\newblock Logistic regression in survival analysis.
\newblock {\em American journal of epidemiology} {\bf 121,} 465--471.

\bibitem[\protect\citeauthoryear{Altman}{Altman}{2007}]{altman2007mixed}
Altman, R.~M. (2007).
\newblock Mixed hidden markov models: an extension of the hidden markov model
  to the longitudinal data setting.
\newblock {\em Journal of the American Statistical Association} {\bf 102,}
  201--210.

\bibitem[\protect\citeauthoryear{Bartolucci and Farcomeni}{Bartolucci and
  Farcomeni}{2015}]{bartolucci2015discrete}
Bartolucci, F. and Farcomeni, A. (2015).
\newblock A discrete time event-history approach to informative drop-out in
  mixed latent markov models with covariates.
\newblock {\em Biometrics} {\bf 71,} 80--89.

\bibitem[\protect\citeauthoryear{Bartolucci and Farcomeni}{Bartolucci and
  Farcomeni}{2019}]{bartolucci2019shared}
Bartolucci, F. and Farcomeni, A. (2019).
\newblock A shared-parameter continuous-time hidden markov and survival model
  for longitudinal data with informative dropout.
\newblock {\em Statistics in medicine} {\bf 38,} 1056--1073.

\bibitem[\protect\citeauthoryear{Baum, Petrie, Soules, and Weiss}{Baum
  et~al.}{1970}]{baum1970maximization}
Baum, L.~E., Petrie, T., Soules, G., and Weiss, N. (1970).
\newblock A maximization technique occurring in the statistical analysis of
  probabilistic functions of markov chains.
\newblock {\em The annals of mathematical statistics} {\bf 41,} 164--171.

\bibitem[\protect\citeauthoryear{Bulla, Bulla, and Nenadi{\'c}}{Bulla
  et~al.}{2010}]{bulla2010hsmm}
Bulla, J., Bulla, I., and Nenadi{\'c}, O. (2010).
\newblock hsmm—an r package for analyzing hidden semi-markov models.
\newblock {\em Computational Statistics \& Data Analysis} {\bf 54,} 611--619.

\bibitem[\protect\citeauthoryear{Bureau, Shiboski, and Hughes}{Bureau
  et~al.}{2003}]{bureau2003applications}
Bureau, A., Shiboski, S., and Hughes, J.~P. (2003).
\newblock Applications of continuous time hidden markov models to the study of
  misclassified disease outcomes.
\newblock {\em Statistics in medicine} {\bf 22,} 441--462.

\bibitem[\protect\citeauthoryear{Callas, Pastides, and Hosmer}{Callas
  et~al.}{1998}]{callas1998empirical}
Callas, P.~W., Pastides, H., and Hosmer, D.~W. (1998).
\newblock Empirical comparisons of proportional hazards, poisson, and logistic
  regression modeling of occupational cohort data.
\newblock {\em American journal of industrial medicine} {\bf 33,} 33--47.

\bibitem[\protect\citeauthoryear{Dempster, Laird, and Rubin}{Dempster
  et~al.}{1977}]{dempster1977maximum}
Dempster, A.~P., Laird, N.~M., and Rubin, D.~B. (1977).
\newblock Maximum likelihood from incomplete data via the em algorithm.
\newblock {\em Journal of the Royal Statistical Society: Series B
  (Methodological)} {\bf 39,} 1--22.

\bibitem[\protect\citeauthoryear{Forney}{Forney}{1973}]{forney1973viterbi}
Forney, G.~D. (1973).
\newblock The viterbi algorithm.
\newblock {\em Proceedings of the IEEE} {\bf 61,} 268--278.

\bibitem[\protect\citeauthoryear{Holsclaw, Greene, Robertson, and
  Smyth}{Holsclaw et~al.}{2017}]{holsclaw2017bayesian}
Holsclaw, T., Greene, A.~M., Robertson, A.~W., and Smyth, P. (2017).
\newblock Bayesian nonhomogeneous markov models via p{\'o}lya-gamma data
  augmentation with applications to rainfall modeling.
\newblock {\em The Annals of Applied Statistics} {\bf 11,} 393--426.

\bibitem[\protect\citeauthoryear{Hothorn}{Hothorn}{2020}]{hothorn2020most}
Hothorn, T. (2020).
\newblock Most likely transformations: The mlt package.
\newblock {\em Journal of Statistical Software} {\bf 92,} v092--i01.

\bibitem[\protect\citeauthoryear{Hothorn, Moest, and Buehlmann}{Hothorn
  et~al.}{2018}]{hothorn2018most}
Hothorn, T., Moest, L., and Buehlmann, P. (2018).
\newblock Most likely transformations.
\newblock {\em Scandinavian Journal of Statistics} {\bf 45,} 110--134.

\bibitem[\protect\citeauthoryear{Hubbard, Lange, Zhang, Salim, Stroud, and
  Inoue}{Hubbard et~al.}{2016}]{hubbard2016using}
Hubbard, R., Lange, J., Zhang, Y., Salim, B., Stroud, J., and Inoue, L. (2016).
\newblock Using semi-markov processes to study timeliness and tests used in the
  diagnostic evaluation of suspected breast cancer.
\newblock {\em Statistics in medicine} {\bf 35,} 4980--4993.

\bibitem[\protect\citeauthoryear{Ingram and Kleinman}{Ingram and
  Kleinman}{1989}]{ingram1989empirical}
Ingram, D.~D. and Kleinman, J.~C. (1989).
\newblock Empirical comparisons of proportional hazards and logistic regression
  models.
\newblock {\em Statistics in medicine} {\bf 8,} 525--538.

\bibitem[\protect\citeauthoryear{Jackson, Sharples, Thompson, Duffy, and
  Couto}{Jackson et~al.}{2003}]{jackson2003multistate}
Jackson, C.~H., Sharples, L.~D., Thompson, S.~G., Duffy, S.~W., and Couto, E.
  (2003).
\newblock Multistate markov models for disease progression with classification
  error.
\newblock {\em Journal of the Royal Statistical Society: Series D (The
  Statistician)} {\bf 52,} 193--209.

\bibitem[\protect\citeauthoryear{Kristensen, Nielsen, Berg, Skaug, and
  Bell}{Kristensen et~al.}{2016}]{kristensen2016tmb}
Kristensen, K., Nielsen, A., Berg, C.~W., Skaug, H., and Bell, B.~M. (2016).
\newblock Tmb: Automatic differentiation and laplace approximation.
\newblock {\em Journal of Statistical Software} {\bf 70,}.

\bibitem[\protect\citeauthoryear{Kr{\'o}l and Saint-Pierre}{Kr{\'o}l and
  Saint-Pierre}{2015}]{krol2015semimarkov}
Kr{\'o}l, A. and Saint-Pierre, P. (2015).
\newblock Semimarkov: An r package for parametric estimation in multi-state
  semi-markov models.
\newblock {\em Journal of Statistical Software} {\bf 66,} 1--16.

\bibitem[\protect\citeauthoryear{Lagona, Jdanov, and Shkolnikova}{Lagona
  et~al.}{2014}]{lagona2014latent}
Lagona, F., Jdanov, D., and Shkolnikova, M. (2014).
\newblock Latent time-varying factors in longitudinal analysis: a linear mixed
  hidden markov model for heart rates.
\newblock {\em Statistics in medicine} {\bf 33,} 4116--4134.

\bibitem[\protect\citeauthoryear{Langrock, Swihart, Caffo, Punjabi, and
  Crainiceanu}{Langrock et~al.}{2013}]{langrock2013combining}
Langrock, R., Swihart, B.~J., Caffo, B.~S., Punjabi, N.~M., and Crainiceanu,
  C.~M. (2013).
\newblock Combining hidden markov models for comparing the dynamics of multiple
  sleep electroencephalograms.
\newblock {\em Statistics in medicine} {\bf 32,} 3342--3356.

\bibitem[\protect\citeauthoryear{Liu, Li, Li, Song, and Rehg}{Liu
  et~al.}{2015}]{liu2015efficient}
Liu, Y.-Y., Li, S., Li, F., Song, L., and Rehg, J.~M. (2015).
\newblock Efficient learning of continuous-time hidden markov models for
  disease progression.
\newblock {\em Advances in neural information processing systems} {\bf 28,}
  3599.

\bibitem[\protect\citeauthoryear{Maruotti and Rocci}{Maruotti and
  Rocci}{2012}]{maruotti2012mixed}
Maruotti, A. and Rocci, R. (2012).
\newblock A mixed non-homogeneous hidden markov model for categorical data,
  with application to alcohol consumption.
\newblock {\em Statistics in Medicine} {\bf 31,} 871--886.

\bibitem[\protect\citeauthoryear{McGilchrist and Aisbett}{McGilchrist and
  Aisbett}{1991}]{mcgilchrist1991regression}
McGilchrist, C. and Aisbett, C. (1991).
\newblock Regression with frailty in survival analysis.
\newblock {\em Biometrics} pages 461--466.

\bibitem[\protect\citeauthoryear{Monk, Kupfer, Frank, and Ritenour}{Monk
  et~al.}{1991}]{monk1991social}
Monk, T.~H., Kupfer, D.~J., Frank, E., and Ritenour, A.~M. (1991).
\newblock The social rhythm metric (srm): measuring daily social rhythms over
  12 weeks.
\newblock {\em Psychiatry research} {\bf 36,} 195--207.

\bibitem[\protect\citeauthoryear{Monk, Flaherty, Frank, Hoskinson, and
  Kupfer}{Monk et~al.}{1990}]{monk1990social}
Monk, T.~K., Flaherty, J.~F., Frank, E., Hoskinson, K., and Kupfer, D.~J.
  (1990).
\newblock The social rhythm metric: An instrument to quantify the daily rhythms
  of life.
\newblock {\em Journal of Nervous and Mental Disease} .

\bibitem[\protect\citeauthoryear{Morris, Aeschbach, and Scheer}{Morris
  et~al.}{2012}]{morris2012circadian}
Morris, C.~J., Aeschbach, D., and Scheer, F.~A. (2012).
\newblock Circadian system, sleep and endocrinology.
\newblock {\em Molecular and cellular endocrinology} {\bf 349,} 91--104.

\bibitem[\protect\citeauthoryear{Shinohara, Sun, and Wang}{Shinohara
  et~al.}{2018}]{shinohara2018alternating}
Shinohara, R.~T., Sun, Y., and Wang, M.-C. (2018).
\newblock Alternating event processes during lifetimes: population dynamics and
  statistical inference.
\newblock {\em Lifetime data analysis} {\bf 24,} 110--125.

\bibitem[\protect\citeauthoryear{Stoner and Economou}{Stoner and
  Economou}{2020}]{stoner2020advanced}
Stoner, O. and Economou, T. (2020).
\newblock An advanced hidden markov model for hourly rainfall time series.
\newblock {\em Computational Statistics \& Data Analysis} {\bf 152,} 107045.

\bibitem[\protect\citeauthoryear{Tam{\'a}si and Hothorn}{Tam{\'a}si and
  Hothorn}{2021}]{tamasi2021tramme}
Tam{\'a}si, B. and Hothorn, T. (2021).
\newblock tramme: Mixed-effects transformation models using template model
  builder.

\bibitem[\protect\citeauthoryear{Therneau and Lumley}{Therneau and
  Lumley}{2015}]{therneau2015package}
Therneau, T.~M. and Lumley, T. (2015).
\newblock Package ‘survival’.
\newblock {\em R Top Doc} {\bf 128,} 28--33.

\bibitem[\protect\citeauthoryear{Thompson~Jr}{Thompson~Jr}{1977}]{thompson1977treatment}
Thompson~Jr, W. (1977).
\newblock On the treatment of grouped observations in life studies.
\newblock {\em Biometrics} pages 463--470.

\bibitem[\protect\citeauthoryear{Torous, Kiang, Lorme, and Onnela}{Torous
  et~al.}{2016}]{torous2016new}
Torous, J., Kiang, M.~V., Lorme, J., and Onnela, J.-P. (2016).
\newblock New tools for new research in psychiatry: a scalable and customizable
  platform to empower data driven smartphone research.
\newblock {\em JMIR mental health} {\bf 3,} e16.

\bibitem[\protect\citeauthoryear{Viterbi}{Viterbi}{1967}]{viterbi1967error}
Viterbi, A. (1967).
\newblock Error bounds for convolutional codes and an asymptotically optimum
  decoding algorithm.
\newblock {\em IEEE transactions on Information Theory} {\bf 13,} 260--269.

\bibitem[\protect\citeauthoryear{Wang, He, and Schaubel}{Wang
  et~al.}{2020}]{wang2020penalized}
Wang, L., He, K., and Schaubel, D.~E. (2020).
\newblock Penalized survival models for the analysis of alternating recurrent
  event data.
\newblock {\em Biometrics} {\bf 76,} 448--459.

\bibitem[\protect\citeauthoryear{Xia, Barnett, Tapera, Adebimpe, Baker,
  Bassett, Brotman, Calkins, Cui, Leibenluft, et~al\mbox{.}}{Xia
  et~al.}{2022}]{xia2022mobile}
Xia, C.~H., Barnett, I., Tapera, T.~M., Adebimpe, A., Baker, J.~T., Bassett,
  D.~S., Brotman, M.~A., Calkins, M.~E., Cui, Z., Leibenluft, E., et~al.
  (2022).
\newblock Mobile footprinting: linking individual distinctiveness in mobility
  patterns to mood, sleep, and brain functional connectivity.
\newblock {\em Neuropsychopharmacology} pages 1--10.

\end{thebibliography}



\pagebreak

\section*{Supporting Information}
Additional supporting information may be found online in the Supporting Information section at the end of the article.


\section*{Figures and Tables}

\begin{table}
\caption{{Simulation: Case Parameters and Accuracy for Competing Methods}.  Mean accuracy and empirical standard errors based on 500 replicates. Each replicate had $I=50$ individuals with $n_{i}=25$ state transitions. The surivial models were simulate with Algorithm 1 and logistic models were simulate with Algorithm 2 from the Web Appendix. Cases 1.1--1.3 looked at low incidence rate and large distributional difference; Cases 2.1--2.3 looked at high incidence rate and large distributional difference; Cases 3.1--3.3 looked at low incidence rate and small distributional difference; and Cases 4.1--4.3 looked at high incidence rate and small distributional difference. }
\begin{center}
\addtolength{\leftskip} {-2cm}
\addtolength{\rightskip}{-2cm}
\footnotesize
\begin{tabular}{ |ccc||cccc|  }
 \hline
 & & & \multicolumn{4}{c|}{Methods: mean accuracy (SE)} \\
 \cline{4-7}
 Case & Simulation & Parameters & PMM & DT-HMM & CT-HMM & PH-HMM \\
 \hline
 1.1 & 
  $\begin{array}{l}
 \text{survival} \\
 h_{\text{max}} = 10
\end{array}$ 
 & 
 $\begin{array}{lll}
 \mu_1=10 & \beta_{10} = -3 & \beta_{11} = -1 \\
 \mu_2=1 & \beta_{20} = -3 & \beta_{21} = 1
\end{array}$ 
 & 0.9851(0.0035) & 0.9904(0.0026) & 0.9835(0.0035) & 0.9842(0.0037) \\
 \hline
 1.2 & 
  $\begin{array}{l}
 \text{survival} \\
 h_{\text{max}} = 1
\end{array}$ 
 & 
 $\begin{array}{lll}
 \mu_1=10 & \beta_{10} = -3 & \beta_{11} = -1 \\
 \mu_2=1 & \beta_{20} = -3 & \beta_{21} = 1
\end{array}$ 
 & 0.9847(0.0215) & 0.9985(0.0011) & 0.9982(0.0012) & 0.9984(0.0011) \\
 \hline
 1.3 & 
  $\begin{array}{l}
 \text{logistic}
\end{array}$ 
 & 
 $\begin{array}{lll}
 \mu_1=10 & \beta_{10} = -3 & \beta_{11} = -1 \\
 \mu_2=1 & \beta_{20} = -3 & \beta_{21} = 1
\end{array}$ 
 & 0.9852(0.0034) & 0.9972(0.0015) & 0.9972(0.0015) & 0.9972(0.0015) \\
\hline \\ \hline
 2.1 & 
  $\begin{array}{l}
 \text{survival} \\
 h_{\text{max}} = 10
\end{array}$ 
 & 
  $\begin{array}{lll}
 \mu_1=10 & \beta_{10} = -2 & \beta_{11} = -5 \\
 \mu_2=1 & \beta_{20} = -2 & \beta_{21} = 5
\end{array}$ 
 & 0.9850(0.0035) & 0.9943(0.0021) & 0.9790(0.0048) & 0.9908(0.0026) \\
 \hline
 2.2 & 
  $\begin{array}{l}
 \text{survival} \\
 h_{\text{max}} = 1
\end{array}$ 
 & 
  $\begin{array}{lll}
 \mu_1=10 & \beta_{10} = -2 & \beta_{11} = -5 \\
 \mu_2=1 & \beta_{20} = -2 & \beta_{21} = 5
\end{array}$ 
 & 0.9856(0.0037) & 0.9980(0.0012) & 0.9974(0.0013) & 0.9978(0.0014) \\
 \hline
 2.3 & 
  $\begin{array}{l}
 \text{logistic} 
\end{array}$ 
 & 
  $\begin{array}{lll}
 \mu_1=10 & \beta_{10} = -2 & \beta_{11} = -5 \\
 \mu_2=1 & \beta_{20} = -2 & \beta_{21} = 5
\end{array}$ 
 & 0.9850(0.0036) & 0.9968(0.0015) & 0.9967(0.0016) & 0.9967(0.0016) \\
\hline \\ \hline
 3.1 & 
  $\begin{array}{l}
 \text{survival} \\
 h_{\text{max}} = 10
\end{array}$ 
 & 
  $\begin{array}{lll}
 \mu_1=5 & \beta_{10} = -3 & \beta_{11} = -1 \\
 \mu_2=1 & \beta_{20} = -3 & \beta_{21} = 1
\end{array}$ 
 & 0.8973(0.0085) & 0.9255(0.0083) & 0.8953(0.0100) & 0.8716(0.0161) \\
 \hline
 3.2 & 
  $\begin{array}{l}
 \text{survival} \\
 h_{\text{max}} = 1
\end{array}$ 
 & 
  $\begin{array}{lll}
 \mu_1=5 & \beta_{10} = -3 & \beta_{11} = -1 \\
 \mu_2=1 & \beta_{20} = -3 & \beta_{21} = 1
\end{array}$ 
 & 0.8966(0.0242) & 0.9879(0.0042) & 0.9865(0.0043) & 0.9820(0.0067) \\
 \hline
 3.3 & 
  $\begin{array}{l}
 \text{logistic} 
\end{array}$ 
 & 
  $\begin{array}{lll}
 \mu_1=5 & \beta_{10} = -3 & \beta_{11} = -1 \\
 \mu_2=1 & \beta_{20} = -3 & \beta_{21} = 1
\end{array}$ 
 & 0.8964(0.0205) & 0.9769(0.0051) & 0.9770(0.0051) & 0.9770(0.0051) \\
\hline \\ \hline
 4.1 & 
  $\begin{array}{l}
 \text{survival} \\
 h_{\text{max}} = 10
\end{array}$ 
 & 
  $\begin{array}{lll}
 \mu_1=5 & \beta_{10} = -2 & \beta_{11} = -5 \\
 \mu_2=1 & \beta_{20} = -2 & \beta_{21} = 5
\end{array}$ 
 & 0.8967(0.0187) & 0.9638(0.0054) & 0.8027(0.0242) & 0.9033(0.0137) \\
 \hline
 4.2 & 
  $\begin{array}{l}
 \text{survival} \\
 h_{\text{max}} = 1
\end{array}$ 
 & 
  $\begin{array}{lll}
 \mu_1=5 & \beta_{10} = -2 & \beta_{11} = -5 \\
 \mu_2=1 & \beta_{20} = -2 & \beta_{21} = 5
\end{array}$ 
 & 0.8945(0.0299) & 0.9862(0.0039) & 0.9841(0.0043) & 0.9778(0.0067) \\
 \hline
 4.3 & 
  $\begin{array}{l}
 \text{logistic} 
\end{array}$ 
 & 
  $\begin{array}{lll}
 \mu_1=5 & \beta_{10} = -2 & \beta_{11} = -5 \\
 \mu_2=1 & \beta_{20} = -2 & \beta_{21} = 5
\end{array}$ 
 & 0.8982(0.0085) & 0.9770(0.0047) & 0.9756(0.0050) & 0.9742(0.0054) \\
\hline
\end{tabular}
\end{center}
\label{sim}
\end{table}

\begin{table}
\caption{{Simulation: Estimates, Standard Errors and Mean Square Errors for ${\mu}_1$ and $\boldsymbol{\beta}_1$}. Mean parameter estimates (Est.), empirical standard error (SE) and mean squared error based (MSE) on 500 replicates. Cases 1.1--1.3 looked at low incidence rate and large distributional difference; Cases 2.1--2.3 looked at high incidence rate and large distributional difference; Cases 3.1--3.3 looked at low incidence rate and small distributional difference; and Cases 4.1--4.3 looked at high incidence rate and small distributional difference.}
\begin{center}
\addtolength{\leftskip} {-2cm}
\addtolength{\rightskip}{-2cm}
\footnotesize
\begin{tabular}{ |cc||rrrr|rrrr|rrrr| }
 \hline
 Case & Model & $\mu_{1}$ & Est. & SE & MSE & $\beta_{10}$ & Est. & SE & MSE & $\beta_{11}$ & Est. & SE & MSE \\
 \hline
 1.1 
& PMM
& 10 & 10.001 & 0.132 & 0.017 & -3 & -2.918 & 0.104 & 0.017 & -1 & -0.921 & 0.136 & 0.025 \\ 
survival
& DT-HMM
& 10 & 9.999 & 0.129 & 0.016 & -3 & -1.294 & 0.116 & 2.923 & -1 & -1.091 & 0.166 & 0.036 \\ 
$h_{\text{max}}=10$
& CT-HMM
& 10 & 9.904 & 0.132 & 0.027 & -3 & -3.075 & 0.107 & 0.017 & -1 & -1.028 & 0.148 & 0.023 \\ 
& PH-HMM
& 10 & 9.885 & 0.134 & 0.031 & -3 & -3.141 & 0.113 & 0.033 & -1 & -1.072 & 0.153 & 0.029 \\ 
 \hline
 1.2
& PMM
& 10 & 9.985 & 0.233 & 0.054 & -3 & -2.113 & 0.261 & 0.855 & -1 & -1.150 & 0.503 & 0.275 \\ 
survival
& DT-HMM
& 10 & 9.993 & 0.123 & 0.015 & -3 & -3.699 & 0.359 & 0.617 & -1 & -1.058 & 0.561 & 0.318 \\ 
$h_{\text{max}}=1$
& CT-HMM
& 10 & 9.987 & 0.124 & 0.015 & -3 & -3.053 & 0.356 & 0.129 & -1 & -1.045 & 0.554 & 0.308 \\ 
& PH-HMM
& 10 & 9.998 & 0.123 & 0.015 & -3 & -2.957 & 0.342 & 0.119 & -1 & -1.032 & 0.538 & 0.290 \\
 \hline
 1.3
& PMM
& 10& 10.000 & 0.130 & 0.017 & -3 & -2.614 & 0.184 & 0.183 & -1 & -0.768 & 0.242 & 0.112 \\ 
logistic
& DT-HMM
& 10 & 10.000 & 0.126 & 0.016 & -3 & -3.013 & 0.217 & 0.047 & -1 & -1.025 & 0.301 & 0.091 \\ 
& CT-HMM
& 10 & 9.999 & 0.126 & 0.016 & -3 & -3.077 & 0.207 & 0.049 & -1 & -0.966 & 0.281 & 0.080 \\ 
& PH-HMM
& 10 & 9.999 & 0.126 & 0.016 & -3 & -3.077 & 0.207 & 0.049 & -1 & -0.966 & 0.281 & 0.080 \\ 
 \hline
 2.1
& PMM
& 10 & 9.990 & 0.129 & 0.017 & -2 & -1.995 & 0.126 & 0.016 & -5 & -3.758 & 0.430 & 1.727 \\ 
survival
& DT-HMM
& 10 & 9.985 & 0.125 & 0.016 & -2 & -0.123 & 0.162 & 3.548 & -5 & -5.684 & 0.574 & 0.797 \\ 
$h_{\text{max}}=10$
& CT-HMM
& 10 & 9.826 & 0.149 & 0.053 & -2 & -2.256 & 0.097 & 0.075 & -5 & -3.353 & 0.345 & 2.831 \\ 
& PH-HMM
& 10 & 9.952 & 0.129 & 0.019 & -2 & -2.129 & 0.106 & 0.028 & -5 & -4.371 & 0.326 & 0.501 \\ 
 \hline
 2.2
& PMM
& 10 & 10.001 & 0.244 & 0.059 & -2 & -1.508 & 0.240 & 0.300 & -5 & -2.759 & 0.523 & 5.296 \\ 
survival
& DT-HMM
& 10 & 10.006 & 0.136 & 0.018 & -2 & -2.679 & 0.395 & 0.618 & -5 & -5.605 & 1.420 & 2.378 \\ 
$h_{\text{max}}=1$
& CT-HMM
& 10 & 9.994 & 0.136 & 0.018 & -2 & -2.135 & 0.380 & 0.162 & -5 & -5.465 & 1.069 & 1.356 \\ 
& PH-HMM
& 10 & 10.013 & 0.136 & 0.019 & -2 & -1.935 & 0.324 & 0.109 & -5 & -4.807 & 0.896 & 0.838 \\ 
 \hline
 2.3
& PMM
& 10 & 9.998 & 0.130 & 0.017 & -2 & -2.016 & 0.159 & 0.025 & -5 & -2.619 & 0.257 & 5.736 \\ 
logistic
& DT-HMM
& 10 & 9.996 & 0.128 & 0.016 & -2 & -2.016 & 0.238 & 0.057 & -5 & -5.173 & 0.722 & 0.550 \\ 
& CT-HMM
& 10 & 9.995 & 0.128 & 0.016 & -2 & -2.348 & 0.166 & 0.149 & -5 & -3.448 & 0.314 & 2.508 \\ 
& PH-HMM
& 10 & 9.997 & 0.128 & 0.016 & -2 & -2.335 & 0.165 & 0.139 & -5 & -3.404 & 0.308 & 2.641 \\ 
 \hline
 3.1
& PMM
& 5 & 5.004 & 0.158 & 0.025 & -3 & -2.509 & 0.073 & 0.247 & -1 & -0.571 & 0.096 & 0.194 \\ 
survival
& DT-HMM
& 5 & 5.016 & 0.114 & 0.013 & -3 & -1.298 & 0.153 & 2.921 & -1 & -1.085 & 0.210 & 0.051 \\ 
$h_{\text{max}}=10$
& CT-HMM
& 5 & 4.819 & 0.127 & 0.049 & -3 & -3.422 & 0.146 & 0.200 & -1 & -1.092 & 0.198 & 0.047 \\ 
& PH-HMM
& 5 & 4.633 & 0.156 & 0.159 & -3 & -4.125 & 0.249 & 1.328 & -1 & -1.414 & 0.309 & 0.267 \\ 
 \hline
 3.2
& PMM
& 5 & 4.997 & 0.215 & 0.046 & -3 & -0.693 & 0.139 & 5.343 & -1 & -0.747 & 0.192 & 0.101 \\ 
survival
& DT-HMM
& 5 & 5.004 & 0.093 & 0.009 & -3 & -3.616 & 0.447 & 0.579 & -1 & -1.146 & 0.631 & 0.419 \\ 
$h_{\text{max}}=1$
& CT-HMM
& 5 & 4.991 & 0.095 & 0.009 & -3 & -3.113 & 0.443 & 0.209 & -1 & -1.111 & 0.606 & 0.378 \\ 
& PH-HMM
& 5 & 5.060 & 0.101 & 0.014 & -3 & -2.265 & 0.401 & 0.701 & -1 & -0.877 & 0.714 & 0.523 \\ 
 \hline
 3.3
& PMM
& 5 & 4.995 & 0.216 & 0.046 & -3 & -1.463 & 0.130 & 2.379 & -1 & -0.382 & 0.123 & 0.396 \\ 
logistic
& DT-HMM
& 5 & 5.007 & 0.098 & 0.010 & -3 & -3.009 & 0.236 & 0.056 & -1 & -1.038 & 0.317 & 0.102 \\ 
& CT-HMM
& 5 & 5.004 & 0.098 & 0.010 & -3 & -3.098 & 0.223 & 0.059 & -1 & -0.990 & 0.292 & 0.085 \\ 
& PH-HMM
& 5 & 5.004 & 0.097 & 0.009 & -3 & -3.097 & 0.223 & 0.059 & -1 & -0.985 & 0.291 & 0.085 \\ 
 \hline
 4.1
& PMM
& 5 & 5.003 & 0.182 & 0.033 & -2 & -1.862 & 0.101 & 0.029 & -5 & -2.085 & 0.173 & 8.529 \\ 
survival
& DT-HMM
& 5 & 5.012 & 0.093 & 0.009 & -2 & -0.143 & 0.235 & 3.502 & -5 & -5.801 & 0.998 & 1.635 \\ 
$h_{\text{max}}=10$
& CT-HMM
& 5 & 4.307 & 0.207 & 0.523 & -2 & -3.215 & 0.220 & 1.525 & -5 & -1.716 & 0.127 & 10.802 \\ 
& PH-HMM
& 5 & 4.865 & 0.136 & 0.037 & -2 & -2.952 & 0.141 & 0.927 & -5 & -2.806 & 0.219 & 4.860 \\ 
 \hline
 4.2
& PMM
& 5 & 4.985 & 0.217 & 0.047 & -2 & -0.481 & 0.140 & 2.326 & -5 & -1.233 & 0.167 & 14.215 \\ 
survival
& DT-HMM
& 5 & 5.013 & 0.090 & 0.008 & -2 & -2.521 & 0.624 & 0.659 & -5 & -5.764 & 2.183 & 5.340 \\ 
$h_{\text{max}}=1$
& CT-HMM
& 5 & 4.994 & 0.090 & 0.008 & -2 & -2.213 & 0.456 & 0.253 & -5 & -5.074 & 1.228 & 1.510 \\ 
& PH-HMM
& 5 & 5.088 & 0.099 & 0.018 & -2 & -1.229 & 0.246 & 0.655 & -5 & -2.477 & 0.413 & 6.534 \\ 
 \hline
 4.3
& PMM
& 5 & 5.010 & 0.154 & 0.024 & -2 & -1.162 & 0.066 & 0.707 & -5 & -1.216 & 0.097 & 14.327 \\ 
logistic
& DT-HMM
& 5 & 5.018 & 0.093 & 0.009 & -2 & -1.973 & 0.319 & 0.102 & -5 & -5.239 & 0.978 & 1.012 \\ 
& CT-HMM
& 5 & 5.015 & 0.091 & 0.009 & -2 & -2.400 & 0.185 & 0.194 & -5 & -3.270 & 0.289 & 3.078 \\ 
& PH-HMM
& 5 & 5.024 & 0.091 & 0.009 & -2 & -2.338 & 0.179 & 0.146 & -5 & -3.033 & 0.273 & 3.945 \\ 
\hline
\end{tabular}
\end{center}
\label{sim_beta}
\end{table}

\begin{table}
\caption{{Simulation: Estimates, Standard Errors and Mean Square Errors for ${\mu}_2$ and $\boldsymbol{\beta}_2$}. Mean parameter estimates, empirical standard error and mean squared error based on 500 replicates. Cases 1.1--1.3 looked at low incidence rate and large distributional difference; Cases 2.1--2.3 looked at high incidence rate and large distributional difference; Cases 3.1--3.3 looked at low incidence rate and small distributional difference; and Cases 4.1--4.3 looked at high incidence rate and small distributional difference.}
\begin{center}
\addtolength{\leftskip} {-2cm}
\addtolength{\rightskip}{-2cm}
\footnotesize
\begin{tabular}{ |cc||rrrr|rrrr|rrrr| }
 \hline
 Case & Model & $\mu_{2}$ & Est. & SE & MSE & $\beta_{20}$ & Est. & SE & MSE & $\beta_{21}$ & Est. & SE & MSE \\
 \hline
 1.1 
& PMM
& 1 & 1.001 & 0.043 & 0.002 & -3 & -2.904 & 0.089 & 0.017 & 1 & 0.909 & 0.125 & 0.024 \\ 
survival
& DT-HMM
& 1 & 0.998 & 0.040 & 0.002 & -3 & -1.291 & 0.105 & 2.932 & 1 & 1.083 & 0.157 & 0.032 \\ 
$h_{\text{max}}=10$
& CT-HMM
& 1 & 1.001 & 0.043 & 0.002 & -3 & -3.059 & 0.095 & 0.013 & 1 & 1.027 & 0.142 & 0.021 \\ 
& PH-HMM
& 1 & 1.004 & 0.043 & 0.002 & -3 & -3.120 & 0.099 & 0.024 & 1 & 1.074 & 0.148 & 0.027 \\ 
 \hline
 1.2
& PMM
& 1 & 1.006 & 0.214 & 0.046 & -3 & -2.140 & 0.297 & 0.828 & 1 & 1.130 & 0.468 & 0.236 \\ 
survival
& DT-HMM
& 1 & 0.994 & 0.038 & 0.001 & -3 & -3.735 & 0.382 & 0.686 & 1 & 1.062 & 0.564 & 0.322 \\ 
$h_{\text{max}}=1$
& CT-HMM
& 1 & 0.995 & 0.038 & 0.001 & -3 & -3.081 & 0.376 & 0.148 & 1 & 1.037 & 0.554 & 0.307 \\ 
& PH-HMM
& 1 & 0.994 & 0.038 & 0.001 & -3 & -2.986 & 0.363 & 0.132 & 1 & 1.088 & 0.551 & 0.310 \\ 
 \hline
 1.3
& PMM
& 1 & 1.001 & 0.043 & 0.002 & -3 & -2.595 & 0.170 & 0.193 & 1 & 0.727 & 0.216 & 0.121 \\ 
logistic
& DT-HMM
& 1 & 0.999 & 0.040 & 0.002 & -3 & -3.009 & 0.207 & 0.043 & 1 & 1.001 & 0.293 & 0.086 \\ 
& CT-HMM
& 1 & 0.999 & 0.040 & 0.002 & -3 & -3.073 & 0.197 & 0.044 & 1 & 0.944 & 0.275 & 0.078 \\ 
& PH-HMM
& 1 & 0.999 & 0.040 & 0.002 & -3 & -3.073 & 0.197 & 0.044 & 1 & 0.944 & 0.275 & 0.078 \\ 
 \hline
 2.1
& PMM
& 1 & 1.003 & 0.043 & 0.002 & -2 & -1.933 & 0.119 & 0.019 & 5 & 3.936 & 0.497 & 1.379 \\ 
survival
& DT-HMM
& 1 & 1.001 & 0.041 & 0.002 & -2 & -0.119 & 0.169 & 3.566 & 5 & 5.612 & 0.525 & 0.649 \\ 
$h_{\text{max}}=10$
& CT-HMM
& 1 & 0.999 & 0.046 & 0.002 & -2 & -2.112 & 0.101 & 0.023 & 5 & 3.838 & 0.435 & 1.540 \\ 
& PH-HMM
& 1 & 1.002 & 0.041 & 0.002 & -2 & -2.090 & 0.109 & 0.020 & 5 & 4.357 & 0.387 & 0.563 \\ 
 \hline
 2.2
& PMM
& 1 & 1.010 & 0.207 & 0.043 & -2 & -1.514 & 0.250 & 0.299 & 5 & 2.824 & 0.472 & 4.958 \\ 
survival
& DT-HMM
& 1 & 0.998 & 0.039 & 0.002 & -2 & -2.655 & 0.354 & 0.554 & 5 & 5.646 & 1.266 & 2.016 \\ 
$h_{\text{max}}=1$
& CT-HMM
& 1 & 0.999 & 0.039 & 0.002 & -2 & -2.076 & 0.341 & 0.122 & 5 & 5.242 & 0.975 & 1.006 \\ 
& PH-HMM
& 1 & 0.998 & 0.039 & 0.002 & -2 & -1.962 & 0.287 & 0.083 & 5 & 4.740 & 0.734 & 0.605 \\ 
 \hline
 2.3
& PMM
& 1 & 1.000 & 0.044 & 0.002 & -2 & -1.933 & 0.146 & 0.026 & 5 & 2.508 & 0.252 & 6.274 \\ 
logistic
& DT-HMM
& 1 & 0.998 & 0.039 & 0.001 & -2 & -2.024 & 0.240 & 0.058 & 5 & 5.198 & 0.724 & 0.562 \\ 
& CT-HMM
& 1 & 0.998 & 0.039 & 0.001 & -2 & -2.352 & 0.166 & 0.152 & 5 & 3.456 & 0.320 & 2.487 \\ 
& PH-HMM
& 1 & 0.998 & 0.039 & 0.001 & -2 & -2.347 & 0.165 & 0.147 & 5 & 3.419 & 0.309 & 2.596 \\ 
 \hline
 3.1
& PMM
& 1 & 1.009 & 0.111 & 0.012 & -3 & -2.615 & 0.076 & 0.154 & 1 & 0.611 & 0.095 & 0.161 \\ 
survival
& DT-HMM
& 1 & 0.979 & 0.053 & 0.003 & -3 & -1.298 & 0.148 & 2.918 & 1 & 1.107 & 0.209 & 0.055 \\ 
$h_{\text{max}}=10$
& CT-HMM
& 1 & 1.011 & 0.065 & 0.004 & -3 & -3.351 & 0.149 & 0.145 & 1 & 1.127 & 0.198 & 0.055 \\ 
& PH-HMM
& 1 & 1.069 & 0.085 & 0.012 & -3 & -4.057 & 0.308 & 1.213 & 1 & 1.633 & 0.379 & 0.543 \\ 
 \hline
 3.2
& PMM
& 1 & 1.024 & 0.205 & 0.043 & -3 & -0.838 & 0.170 & 4.701 & 1 & 1.026 & 0.184 & 0.035 \\ 
survival
& DT-HMM
& 1 & 0.986 & 0.042 & 0.002 & -3 & -3.779 & 0.697 & 1.091 & 1 & 1.192 & 0.880 & 0.810 \\ 
$h_{\text{max}}=1$
& CT-HMM
& 1 & 0.990 & 0.043 & 0.002 & -3 & -3.274 & 0.696 & 0.558 & 1 & 1.158 & 0.877 & 0.793 \\ 
& PH-HMM
& 1 & 0.977 & 0.043 & 0.002 & -3 & -2.396 & 0.988 & 1.339 & 1 & 1.390 & 1.195 & 1.577 \\ 
 \hline
 3.3
& PMM
& 1 & 1.017 & 0.155 & 0.024 & -3 & -1.581 & 0.228 & 2.064 & 1 & 0.509 & 0.149 & 0.263 \\ 
logistic
& DT-HMM
& 1 & 0.984 & 0.043 & 0.002 & -3 & -3.060 & 0.248 & 0.065 & 1 & 1.083 & 0.350 & 0.129 \\ 
& CT-HMM
& 1 & 0.985 & 0.043 & 0.002 & -3 & -3.148 & 0.235 & 0.077 & 1 & 1.030 & 0.323 & 0.105 \\ 
& PH-HMM
& 1 & 0.985 & 0.043 & 0.002 & -3 & -3.146 & 0.234 & 0.076 & 1 & 1.025 & 0.322 & 0.104 \\ 
 \hline
 4.1
& PMM
& 1 & 1.008 & 0.141 & 0.020 & -2 & -2.123 & 0.207 & 0.058 & 5 & 2.038 & 0.151 & 8.794 \\ 
survival
& DT-HMM
& 1 & 0.993 & 0.044 & 0.002 & -2 & -0.143 & 0.225 & 3.499 & 5 & 5.699 & 0.819 & 1.158 \\ 
$h_{\text{max}}=10$
& CT-HMM
& 1 & 0.897 & 0.107 & 0.022 & -2 & -2.250 & 0.157 & 0.087 & 5 & 1.604 & 0.177 & 11.564 \\ 
& PH-HMM
& 1 & 0.992 & 0.064 & 0.004 & -2 & -2.798 & 0.130 & 0.654 & 5 & 2.921 & 0.246 & 4.383 \\ 
 \hline
 4.2
& PMM
& 1 & 1.013 & 0.199 & 0.040 & -2 & -0.690 & 0.291 & 1.801 & 5 & 1.497 & 0.162 & 12.300 \\ 
survival
& DT-HMM
& 1 & 0.986 & 0.043 & 0.002 & -2 & -2.645 & 0.531 & 0.697 & 5 & 6.051 & 2.109 & 5.543 \\ 
$h_{\text{max}}=1$
& CT-HMM
& 1 & 0.990 & 0.043 & 0.002 & -2 & -2.272 & 0.379 & 0.217 & 5 & 4.916 & 1.030 & 1.065 \\ 
& PH-HMM
& 1 & 0.973 & 0.044 & 0.003 & -2 & -1.447 & 0.240 & 0.363 & 5 & 2.745 & 0.341 & 5.201 \\ 
 \hline
 4.3
& PMM
& 1 & 1.007 & 0.112 & 0.013 & -2 & -1.344 & 0.075 & 0.435 & 5 & 1.443 & 0.103 & 12.664 \\ 
logistic
& DT-HMM
& 1 & 0.987 & 0.041 & 0.002 & -2 & -2.001 & 0.341 & 0.116 & 5 & 5.295 & 1.026 & 1.137 \\ 
& CT-HMM
& 1 & 0.984 & 0.041 & 0.002 & -2 & -2.424 & 0.190 & 0.216 & 5 & 3.267 & 0.308 & 3.097 \\ 
& PH-HMM
& 1 & 0.980 & 0.041 & 0.002 & -2 & -2.384 & 0.183 & 0.181 & 5 & 3.040 & 0.276 & 3.919 \\ 
\hline
\end{tabular}
\end{center}
\label{sim_etc}
\end{table}

\begin{table}
\caption{{Population HMM: Parameter Estimates for Discrete-Time Setting}. Parameter estimates using EM algorithm and asymptotic standard errors. Regression for state transitions are given as $\eta_s(t_{ij}) = \beta_{s0} + \beta_{s1} x(t_{ij}) + \beta_{s2}x(t_{ij}) \mathrm{sex} + \beta_{s3}x(t_{ij}) \mathrm{OS} + \mathbf{z}^\top (t_{ij}) \mathbf{b}_{s}$ where $\mathbf{b}_s \sim \mathrm{N}(\mathbf{0}_{41}, \sigma^2_s\mathbf{I}_{41} )$ are individual specific random intercepts, Android devices and males serve as baseline. }
\begin{center}
\addtolength{\leftskip} {-2cm}
\addtolength{\rightskip}{-2cm}
\small
\begin{tabular}{ |cc||cccc|  }
 \hline
& & \multicolumn{4}{c|}{Methods: Estimate (SE)} \\
 \cline{3-6}
 Transition & Parameters & PMM & DT-HMM & CT-HMM & PH-HMM \\
 \hline
& $\beta_{10}$ & 2.8973(0.8384) & 10.8254(1.4074) & 9.0154(1.2162) & 8.5893(1.2036) \\ 
Active & $\beta_{11}$ & -0.4182(0.0850) & -1.2431(0.1428) & -1.0828(0.1232) & -1.0395(0.1219) \\ 
to & $\beta_{12}$ & 0.0288(0.0121) & 0.0379(0.0184) & 0.0257(0.0126) & 0.0257(0.0126) \\ 
Rest & $\beta_{13}$ & 0.0070(0.0124) & 0.0052(0.0189) & 0.0054(0.0130) & 0.0050(0.0129) \\ 
& $\sigma^2_{1}$ & 0.1099 & 0.2585 & 0.1171 & 0.1160 \\ 
 \hline
& $\beta_{20}$ & -6.6357(0.3846) & -16.2898(1.3782) & -6.0404(0.4941) & -5.9631(0.5029) \\ 
Rest & $\beta_{21}$ & 0.5359(0.0405) & 1.5280(0.1401) & 0.4671(0.0512) & 0.4599(0.0521) \\ 
to & $\beta_{22}$ & -0.0278(0.0190) & -0.0441(0.0203) & -0.0357(0.0183) & -0.0355(0.0183) \\ 
Active & $\beta_{23}$ & -0.0431(0.0191) & -0.0541(0.0208) & -0.0503(0.0187) & -0.0505(0.0188) \\ 
& $\sigma^2_{2}$ & 0.2802 & 0.3210 & 0.2631 & 0.2634 \\ 
 \hline
& $\mu_1$ & 8.5321 & 8.0866 & 8.0251 & 8.0114 \\
& $\mu_2$ & 0.5238 & 0.4263 & 0.4195 & 0.4153 \\
 \hline
\end{tabular}
\end{center}
\label{HMM_fit}
\end{table}

\begin{table}
\caption{{Population HMM: Parameter Estimates for Heterogeneous Event Time Setting}. Parameter estimates using EM algorithm and asymptotic standard errors. Regression for state transitions are given as $\eta_s(t_{ij}) = \beta_{s0} + \beta_{s1} x(t_{ij}) + \beta_{s2}x(t_{ij}) \mathrm{sex} + \beta_{s3}x(t_{ij}) \mathrm{OS} + \mathbf{z}^\top (t_{ij}) \mathbf{b}_{s}$ where $\mathbf{b}_s \sim \mathrm{N}(\mathbf{0}_{41}, \sigma^2_s\mathbf{I}_{41} )$ are individual specific random intercepts, Android devices and males serve as baseline. }
\begin{center}
\addtolength{\leftskip} {-2cm}
\addtolength{\rightskip}{-2cm}
\small
\begin{tabular}{ |cc||cccc|  }
 \hline
 & & \multicolumn{4}{c|}{Methods: Estimate (SE)} \\
 \cline{3-6}
 Transition & Parameters & PMM & DT-HMM & CT-HMM & PH-HMM \\
 \hline
& $\beta_{10}$ & 0.3714(0.7637) & 6.4420(1.2828) & 2.8110(1.1055) & 2.0846(1.0867) \\ 
Active & $\beta_{11}$ & -0.2567(0.0786) & -0.7771(0.1305) & -0.5526(0.1130) & -0.5209(0.1120) \\ 
to & $\beta_{12}$ & 0.0541(0.0214) & 0.0496(0.0231) & 0.0528(0.0233) & 0.0685(0.0285) \\ 
Rest & $\beta_{13}$ & 0.0867(0.0221) & 0.0094(0.0241) & 0.0840(0.0240) & 0.1127(0.0296) \\ 
& $\sigma^2_{1}$ & 0.3554 & 0.3941 & 0.4102 & 0.6132 \\ 
 \hline
& $\beta_{20}$ & -6.4771(0.3775) & -20.1172(1.2551) & -6.9784(0.4544) & -6.8518(0.4673) \\ 
Rest & $\beta_{21}$ & 0.4597(0.0411) & 2.0046(0.1298) & 0.5196(0.0481) & 0.4889(0.0500) \\ 
to & $\beta_{22}$ & -0.0211(0.0235) & -0.0620(0.0294) & -0.0383(0.0217) & -0.0327(0.0241) \\ 
Active & $\beta_{23}$ & 0.0033(0.0240) & -0.0874(0.0309) & -0.0199(0.0225) & -0.0077(0.0252) \\ 
& $\sigma^2_{2}$ & 0.4051 & 0.6207 & 0.3422 & 0.4040 \\ 
 \hline
& $\mu_1$ & 9.3939 & 9.0530 & 8.9522 & 8.9428 \\
& $\mu_2$ & 1.0534 & 0.9579 & 0.9528 & 0.9534 \\
 \hline
\end{tabular}
\end{center}
\label{HMM_fit2}
\end{table}


\label{lastpage}

\end{document}